\documentclass[sn-mathphys]{sn-jnl}
\jyear{2021}%

\theoremstyle{thmstyleone}%
\usepackage{booktabs} 
\usepackage{arydshln}

\theoremstyle{thmstyletwo}%

\theoremstyle{thmstylethree}%

\usepackage{amsmath,amssymb,amsfonts}%
\usepackage{cleveref}
\usepackage{color}
\usepackage{multirow}
\usepackage{graphicx}
\usepackage{comment}
\usepackage{changepage}
\usepackage{wrapfig}
\usepackage{colortbl} 
\usepackage{arydshln} 
\usepackage{amsthm}
\usepackage{bm}
\usepackage{rotating}
\usepackage{amsmath}
\usepackage{mathrsfs}%
\usepackage{textcomp}%
\usepackage{manyfoot}%

\newcommand{\tightdoublehline}{\noalign{\vskip 0.5ex}\hline\noalign{\vskip 0.5ex}\hline\noalign{\vskip 0.5ex}}
\newcommand{\tightdoublehlineplus}{\noalign{\vskip -0.6ex}\cmidrule(lr){2-11}\noalign{\vskip -0.6ex}\cmidrule(lr){2-11}\noalign{\vskip -0.6ex}}

\begin{document}

\title[ ]{Mamba-driven multi-perspective structural understanding for molecular ground-state conformation prediction}


%\author*[1,2]{\fnm{First} \sur{Author}}\email{iauthor@gmail.com}
%\affil[3]{\orgdiv{Department}, \orgname{Organization}, \orgaddress{\street{Street}, \city{City}, \postcode{610101}, \state{State}, \country{Country}}}

\author[1]{\fnm{Yuxin} \sur{Gou}}\email{gouyuxin@mail.hfut.edu.cn}

\author*[1]{\fnm{Aming} \sur{Wu}}\email{amwu@hfut.edu.cn}

\author[1]{\fnm{Richang} \sur{Hong}}\email{hongrc.hfut@gmail.com}

\author*[1]{\fnm{Meng} \sur{Wang}}\email{eric.mengwang@gmail.com}

\affil[1]{\orgdiv{School of Computer Science and Information Engineering}, \orgname{Hefei University of Technology}, \orgaddress{\city{Hefei}, \state{Anhui Province}, \country{China}}}


\abstract{A comprehensive understanding of molecular structures is important for the prediction of molecular ground-state conformation involving property information. Meanwhile, state space model (e.g., Mamba) has recently emerged as a promising mechanism for long sequence modeling and has achieved remarkable results in various language and vision tasks. However, towards molecular ground-state conformation prediction, exploiting Mamba to understand molecular structure is underexplored. To this end, we strive to design a generic and efficient framework with Mamba to capture critical components. In general, molecular structure could be considered to consist of three elements, i.e., atom types, atom positions, and connections between atoms. Thus, considering the three elements, an approach of Mamba-driven multi-perspective structural understanding (MPSU-Mamba) is proposed to localize molecular ground-state conformation. Particularly, for complex and diverse molecules, three different kinds of dedicated scanning strategies are explored to construct a comprehensive perception of corresponding molecular structures. And a bright-channel guided mechanism is defined to discriminate the critical conformation-related atom information. Experimental results on QM9 and Molecule3D datasets indicate that MPSU-Mamba significantly outperforms existing methods. Furthermore, we observe that for the case of few training samples, MPSU-Mamba still achieves superior performance, demonstrating that our method is indeed beneficial for understanding molecular structures.}




\maketitle
\vspace{-0.2in}
\section{Introduction}\label{sec1}

Molecular ground-state conformation represents the most stable 3D structure with the minimized energy on its potential energy surface~\cite{xu2023gtmgc,kim2025rebind}. This ground-state conformation typically corresponds to multiple important molecular properties, e.g., molecule's physical, chemical, and biological characteristics~\cite{xu2023gtmgc,wang2024wgformer}. As a result, it plays a crucial role in various downstream applications, e.g., molecular property prediction \cite{schutt2021equivariant,ni2023sliced,zaidi2022pre}, drug discovery \cite{zhou2023deep,tang2024survey}, and protein-ligand interactions \cite{pei2023fabind,wang2024structure}.

With the rejuvenation of deep neural networks, in order to accelerate the development of biochemical research, molecular representation learning (MRL) has attracted growing attention and achieved many advances \cite{boulougouri2024molecular,wan2025multi,bai2023interpretable,wang2023multitask,fang2023knowledge,zhuang2023learning,yang2022learning}. Particularly, molecules are usually represented as graphs \cite{kipf2016semi,li2020deepergcn,ying2021transformers,kim2022pure}, where nodes and edges separately depict corresponding atoms and bonds. Taking 2D molecular graphs as input, 3D molecular conformations can be predicted by performing iterative message-passing updates \cite{xu2023gtmgc}. Essentially, through defining proper pairwise distances, bonded and proximate atomic pairs are encouraged to be close in the representational space \cite{xu2023gtmgc,wang2024wgformer}. Though graph-based methods have been evaluated to be effective on multiple datasets \cite{ramakrishnan2014quantum,xu2021molecule3d}, for diverse and complex macromolecules, only utilizing the simple pairwise distance is difficult to accurately capture the structural information that has the lowest possible energy, weakening the prediction performance. The reason may be that graph neural networks usually depend on the weighting methodology for features from neighbouring nodes \cite{dai2025survey}, which could not effectively capture complex relationships of long-sequence atoms.

In general, the molecular ground-state conformation could be considered as a specific structure of a molecule. Obviously, a comprehensive understanding of molecular structure is instrumental in discriminating ground-state conformation. Particularly, atom types, atom positions, and connections between atoms all play important roles in determining the formation of molecular structure \cite{wells2012structural}. Thus, we explore strengthening the understanding of molecular structure from these three perspectives, which improves the recognition performance of molecular ground-state conformation.

Currently, Transformer network has been a popular architecture in various tasks and has been shown to be effective on many datasets \cite{vaswani2017attention,devlin2019bert,carion2020end,khan2022transformers,ye2024differential}. By decomposing the input into a token sequence and stacking multi-layer attention, the task-related token information could be gradually focused, boosting the performance of current tasks. Whereas, as a local component of a molecule, directly disassembling tokens may destroy the complete content of the ground-state conformation. Meanwhile, calculating attention weights may ignore the information of the overall molecular structure, affecting the localization accuracy of the ground-state conformation.

Furthermore, state space model \cite{gu2021efficiently} has recently emerged as a promising mechanism for capturing long-sequence dependencies. As a representative work, Mamba \cite{gu2023mamba,behrouz2024mambamixer,pioro2024moe,wang2024mambabyte,zhu2024vision} incorporates time-varying parameters into state space model and proposes a hardware-aware algorithm to enable efficient training and inference. Though Mamba has achieved remarkable success in various vision and language tasks, towards molecular conformation prediction, employing Mamba to understand molecular structure is rarely explored. The challenge mainly lies in how to design an effective scanning strategy to obtain sufficient information about molecular structures.

To this end, we propose an approach of Mamba-driven multi-perspective structural understanding (MPSU-Mamba) to localize molecular ground-state conformation. Concretely, as shown in Fig. \ref{fig1}, for complex and diverse molecules, three different kinds of dedicated scanning strategies are explored to construct a comprehensive perception from atom type, position, and connection perspectives. Meanwhile, a bright-channel guided mechanism is defined to discriminate the critical conformation-relevant atom information. Experimental results on multiple datasets indicate that MPSU-Mamba significantly outperforms existing methods. Moreover, we also observe that for the case of few training samples, MPSU-Mamba still achieves superior prediction performance, demonstrating not only that capturing molecular structural information could enhance the ability of conformation prediction but also that our method is indeed instrumental in understanding molecular structure.

\section{Results}

\subsection{Overview of MPSU-Mamba Framework}

In this paper, we demonstrate that a comprehensive understanding of molecular structure is beneficial for achieving an accurate prediction of the molecular ground-state conformation. Motivated by the success of Mamba in various language and vision tasks \cite{dai2025orochi,wang2025reasoning,lu2025mamba,kotoge2025evobrain,chharia2025mv}, we explore designing a generic and efficient framework with Mamba to capture critical components from multiple different structural perspectives.
\begin{figure*}[t]
  \centering
  \includegraphics[width=\textwidth]{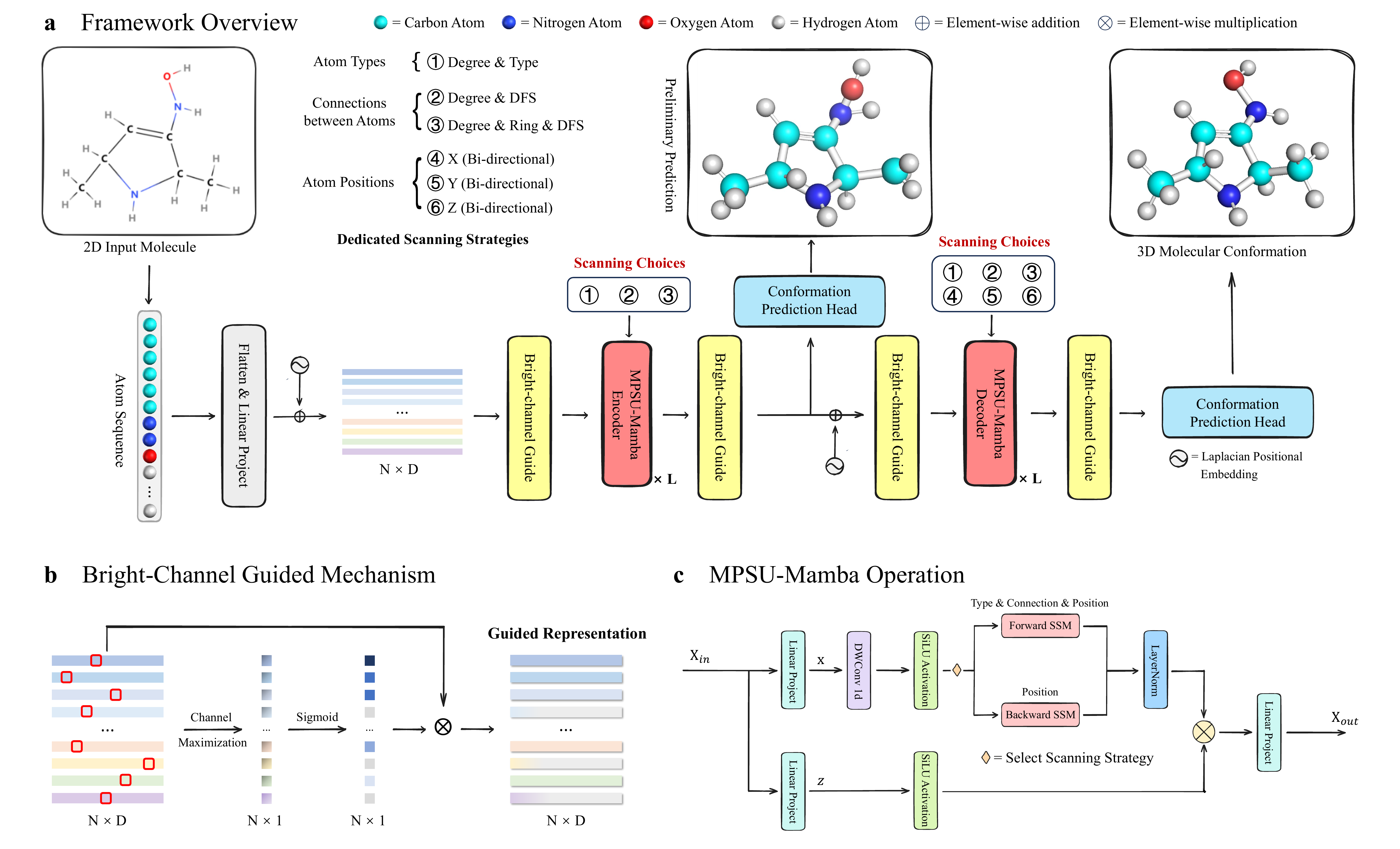}
  \vspace{-0.2in}
  \caption{The overview of the proposed MPSU-Mamba framework. a, Mamba-driven multi-perspective molecular structural understanding for 3D conformation prediction, mainly comprising bright-channel guided operations, MPSU-Mamba encoder, MPSU-Mamba decoder, and 3D conformation prediction head. Particularly, given a 2D input molecule, we flatten it into a sequence of atoms. After obtaining the projected features, a bright-channel guided operation is first performed to focus on critical conformation-aware information. Then, we define multiple Mamba layers to extract the encoding representation. Here, we utilize the scanning strategies based on atom types and connections between atoms. Next, a conformation prediction head is defined to estimate the initial 3D positions. Finally, we still design a multi-layer Mamba as the decoder to achieve the corresponding 3D molecular conformation. Here, the scanning strategies are selected based on atom types, connections between atoms, and atom positions. b, The details of Bright-Channel guided operation. We observe that by stacking multi-layer guided operations, the accuracy of conformation prediction could be improved effectively. c, View of a MPSU-Mamba operation. For type- and connection-based strategies, we only utilize the forward direction. Differently, for the position strategy, the forward and backward directions are used.}
  \label{fig1}
  \vspace{-0.1in}
\end{figure*}
Specifically, as illustrated in Fig. \ref{fig1}a, given a 2D input molecule, we randomly flatten it into a sequence of atoms. A linear projector is utilized to obtain the embedding features. Here, we assume that in a high-dimensional representation space, the maximized channel may involve significant task-relevant information. Therefore, in Fig. \ref{fig1}b, we design a bright-channel guided mechanism to focus on conformation-aware atom characteristics. Through multiple guided operations, the performance of conformation prediction could be improved effectively.

\begin{figure*}[t]
  \centering
  \includegraphics[width=\textwidth]{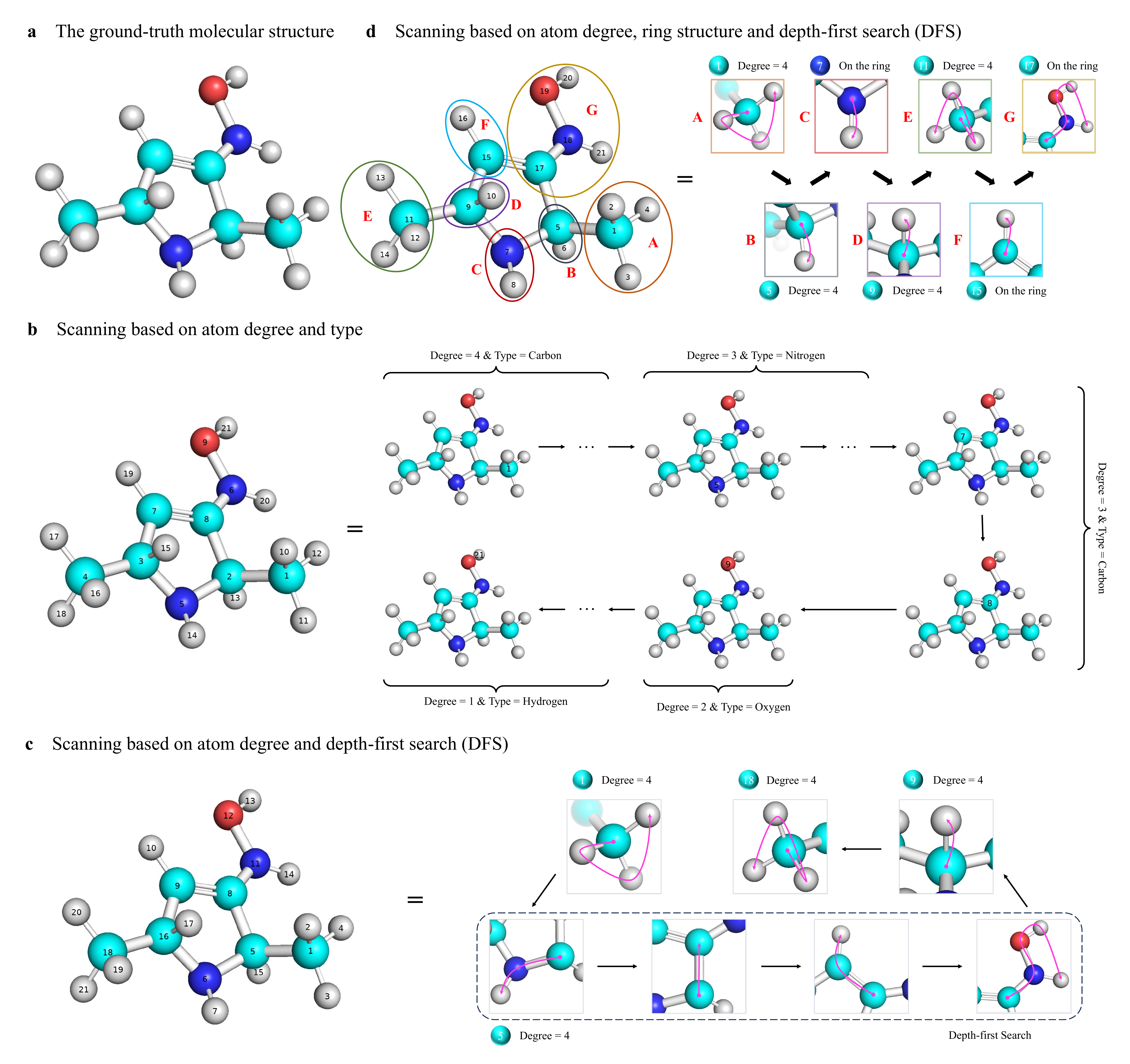}
  \vspace{-0.3in}
  \caption{Visualization examples of different scanning strategies. Here, the number in each atom depicts the corresponding scanning order. To obtain a comprehensive understanding of molecular structures, a series of dedicated scanning mechanisms are defined from two various perspectives, i.e., atom types and connections between atoms. Extensive experimental results on multiple benchmarks demonstrate that these strategies could effectively improve the performance of conformation prediction.}
  \label{fig2}
  \vspace{-0.1in}
\end{figure*}

\begin{figure*}[t]
  \centering
  \includegraphics[width=\textwidth]{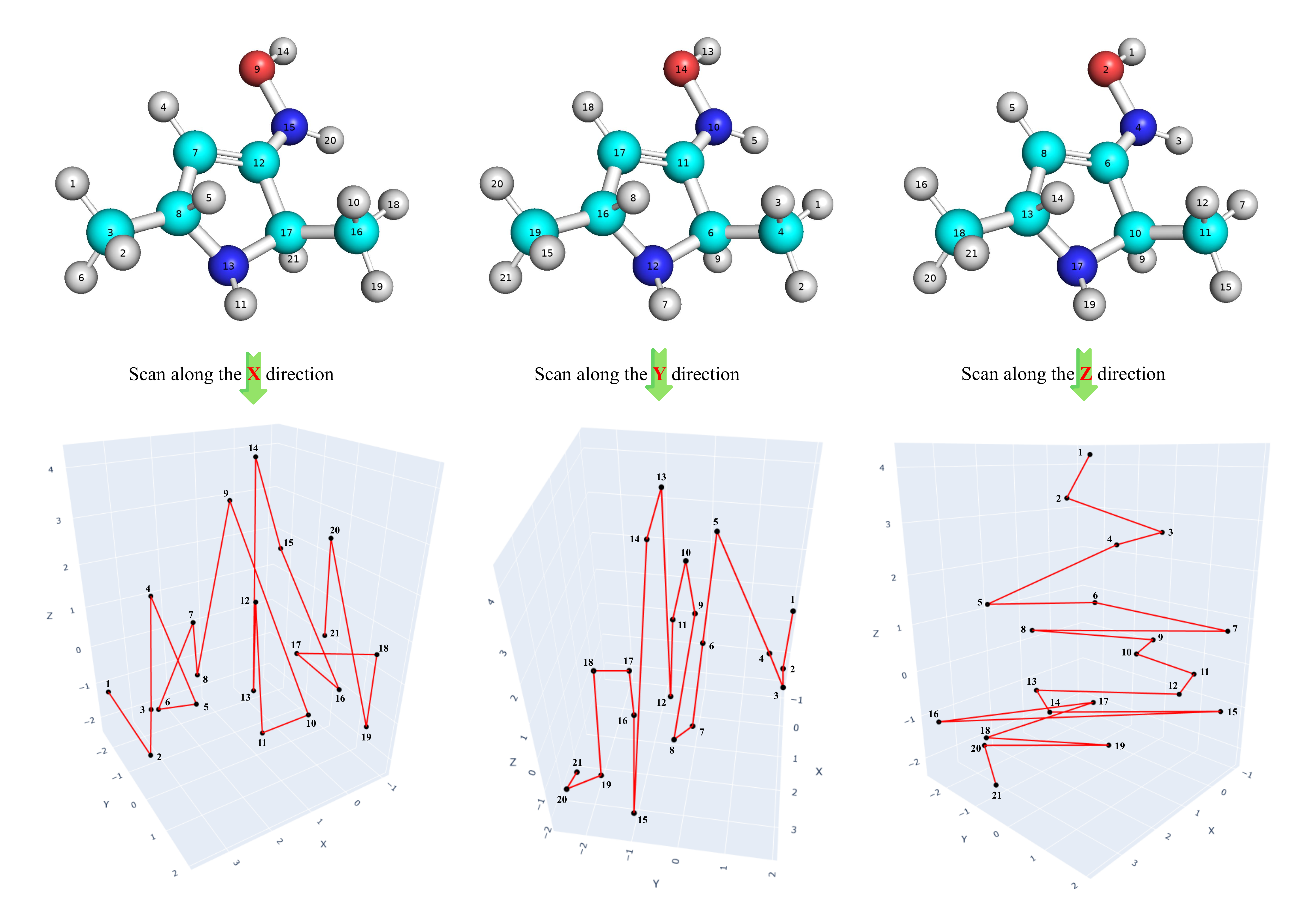}
  \vspace{-0.3in}
  \caption{Scanning based on 3D atom coordinates (Descending order). Since the coordinates are produced based on the encoding representation, using position-based scanning could not only enhance the understanding level but also improve the accuracy of the final prediction.}
  \label{fig3}
  \vspace{-0.1in}
\end{figure*}

Then, we stack multiple Mamba layers to extract the encoding molecular representation. As shown in Fig. \ref{fig1} and \ref{fig2}, we first exploit the scanning strategies based on atom types and connections between atoms. For example, the atoms in the ring are usually more stable. Thus, in the scanning process, we consider the factor of the ring structure. By means of these operations, the encoder could construct initial concepts about the structures of given molecules. Next, based on the encoding representation, a confirmation head is defined to preliminarily predict 3D coordinates. Subsequently, multiple layers of Mamba are utilized to achieve the decoding results. Here, during the decoding process, besides the scanning strategies used in the encoder, we consider the position element. Particularly, as shown in Fig. \ref{fig3}, various scanning orders are separately built based on the X, Y, and Z coordinates. Meanwhile, to further deepen the structure understanding, we also employ the operation of bidirectional scanning. This operation is not only conducive to precisely understanding molecular structures but also instrumental in improving the accuracy of the preliminary predictions through optimization. Finally, we discover that by constructing proper scanning mechanisms, this approach is more effective than simply utilizing Graph and Transformer networks. Furthermore, from the human perspective, different scanning strategies contain various observation views on molecular structures. By combining these perspectives, we can establish a comprehensive perception of complex and diverse molecular structures, which improves the performance of downstream tasks. %More details are provided in the Method section.

\subsection{Performance on Small-scale Organic Molecules}

To demonstrate the effectiveness of our approach, we first verify MPSU-Mamba on small-scale organic molecules. Here, QM9 dataset \cite{ramakrishnan2014quantum,wu2018moleculenet,liao2022equiformer} is utilized for evaluation, which provides geometric, energy, electronic, and thermodynamic properties for nearly 130, 000 organic molecules with 9 heavy atoms, containing the most stable molecular conformation calculated by Density Functional Theory (DFT). Meanwhile, we exploit four metrics, i.e., D-MAE, D-RMSE, C-RMSD, and E-RMSD, to evaluate the experimental results. Table \ref{table1} shows the corresponding performance.

As shown in Fig. \ref{fig1}, the task of ground-state conformation prediction aims to predict the 3D conformation of a molecule in its ground state, $\mathbf{G} \in \mathbb{R}^{\rm N \times 3}$, solely based on its 2D molecular structure, $\mathcal{G} = \{ \mathcal{V}, \mathcal{E} \}$, where ${\rm N}$ indicates the atom number. Existing solutions could be categorized into two types: 2D and 3D methods. Particularly, 2D methods \cite{xu2023gtmgc,kim2025rebind} directly utilize the 2D input and predict corresponding 3D conformations. Differently, WGFormer \cite{wang2024wgformer} first applies the cheminformatic tool RDKit \cite{landrum2013rdkit} to obtain an initial low-quality 3D conformation for the given 2D molecular graph and then predict the 3D ground-state conformation, which can be considered as a 3D-to-3D method. Therefore, the performance of 3D methods may be affected by the quality of initial conformation. In Fig. \ref{fig1}, our method belongs to the 2D-to-3D type, directly predicting 3D conformation based on 2D input.

In Table \ref{table1}, we first compare with four state-of-the-art baselines, including GTMGC \cite{xu2023gtmgc}, REBIN \cite{kim2025rebind}, DiffFormer \cite{ye2024differential}, and CFD \cite{amingcfd}. Particularly, GTMGC, REBIND, and DiffFormer separately present dedicated attention mechanisms and stack multiple attention layers to gradually focus on critical conformation-aware components. And CFD designs an iterative representation mechanism. Compared with biological macromolecules, e.g., proteins, small organic molecules generally possess fewer atoms and simpler structures. However, based on the Mean Absolute Error (MAE) metric, the prediction accuracy is still unsatisfactory, e.g., the error is higher than 20\% \cite{kim2025rebind,xu2023gtmgc}, which indicates the challenge of this task. We can observe that based on the four metrics, our approach significantly outperforms the four compared methods. For example, compared with REBIND \cite{kim2025rebind}, based on D-MAE, D-RMSE, C-RMSD, and E-RMSD metrics, the performance is improved by around 16.6\%, 22.4\%, 15.4\%, and 26.9\%. Meanwhile, our method also outperforms WGFormer \cite{wang2024wgformer}, e.g., the D-MAE performance is boosted by around 13.9\%. Essentially, the goal of attention mechanisms is to assign a different weight to the corresponding atom, which depicts its importance for the overall molecule. Only using the attention mechanism usually pays attention to certain atoms and is difficult to understand the complete structure, which weakens the prediction performance. 

\begin{table*}[htbp]
  \centering
  \caption{The performance comparison of various models on QM9 \cite{ramakrishnan2014quantum,wu2018moleculenet} dataset. Here, we bold the best result and underline the second-best one. `-' represents the unreported result.}
  \vspace{-0.05in}
  \small{
    \resizebox{\textwidth}{!}{
    \begin{tabular}{cccccccccc}
    \toprule
    \multirow{2}[4]{*}{Method} & \multirow{2}[4]{*}{Model} & \multicolumn{4}{c}{Validation} & \multicolumn{4}{c}{Test} \\
    % \cmidrule{3-10}          
    \cmidrule(lr){3-6} \cmidrule(lr){7-10}
    &       & D-MAE $\downarrow$ & D-RMSE $\downarrow$ & C-RMSD $\downarrow$ & E-RMSD $\downarrow$ & D-MAE $\downarrow$ & D-RMSE $\downarrow$ & C-RMSD $\downarrow$ & E-RMSD $\downarrow$ \\
    \midrule
    \multirow{5}[2]{*}{\textbf{3D}} & SE(3)-Transformer \cite{fuchs2020se} & 0.254 & 0.451 & 0.296 & -     & 0.256 & 0.455 & 0.303 & - \\
          & EGNN \cite{satorras2021n} & 0.248 & 0.442 & 0.257 & -     & 0.251 & 0.449 & 0.265 & - \\
          & ConfOpt-TwoAtom \cite{guan2021energy} & 0.245 & 0.439 & 0.245 & -     & 0.248 & 0.444 & 0.254 & - \\
          & ConfOpt-ThreeAtom \cite{guan2021energy} & 0.241 & 0.433 & 0.237 & -     & 0.244 & 0.438 & 0.246 & - \\
          & WGFormer \cite{wang2024wgformer} & \underline{0.223} & \underline{0.416} & \underline{0.198} & -     & 0.227 & \underline{0.422} & \underline{0.206} & - \\
    \tightdoublehline
    \multirow{10}[2]{*}{\textbf{2D}}
          & RDKit DG \cite{landrum2013rdkit} & 0.358 & 0.616 & 0.722 & 1.044 & 0.358 & 0.615 & 0.722 & 1.266 \\
          & RDKit ETKDG \cite{landrum2013rdkit} & 0.355 & 0.621 & 0.691 & 1.048 & 0.355 & 0.621 & 0.689 & 1.120 \\
          & GINE \cite{hu2019strategies}  & 0.357 & 0.673 & 0.685 & 1.703 & 0.357 & 0.669 & 0.693 & 1.696 \\
          & GATv2 \cite{brody2021attentive} & 0.339 & 0.663 & 0.661 & 1.371 & 0.339 & 0.659 & 0.666 & 1.358 \\
          & GPS \cite{rampavsek2022recipe}  & 0.326 & 0.644 & 0.662 & 1.196 & 0.326 & 0.640 & 0.666 & 1.193 \\
          & GTMGC \cite{xu2023gtmgc} & 0.262 & 0.468 & 0.362 & 0.792 & 0.264 & 0.470 & 0.367 & 0.800 \\
          & REBIND \cite{kim2025rebind} & 0.252 & 0.442 & 0.320 & 0.601 & 0.254 & 0.446 & 0.321 & 0.610 \\
          & DiffFormer \cite{ye2024differential}  & 0.247 & 0.442 & 0.308 & \underline{0.571} & 0.249 & 0.445 & 0.313 & \underline{0.581} \\
          & CFD \cite{amingcfd}  & \underline{0.223} & 0.434 & 0.305 & -     & \underline{0.218} & 0.442 & 0.309 & - \\
          & \cellcolor[rgb]{ .851,  .882,  .957}MPSU-Mamba(ours) & \cellcolor[rgb]{ .851,  .882,  .957}\textbf{0.087} & \cellcolor[rgb]{ .851,  .882,  .957}\textbf{0.217} & \cellcolor[rgb]{ .851,  .882,  .957}\textbf{0.164} & \cellcolor[rgb]{ .851,  .882,  .957}\textbf{0.331} & \cellcolor[rgb]{ .851,  .882,  .957}\textbf{0.088} & \cellcolor[rgb]{ .851,  .882,  .957}\textbf{0.222} & \cellcolor[rgb]{ .851,  .882,  .957}\textbf{0.167} & \cellcolor[rgb]{ .851,  .882,  .957}\textbf{0.341} \\
    \bottomrule
    \end{tabular}%
    }
  }
  \label{table1}%
  \vspace{-0.1in}
\end{table*}

\begin{table*}[htbp]
  \centering
  \caption{The performance comparison of conformation prediction on Molecule3D~\cite{xu2021molecule3d} dataset based on Random and Scaffold splits.}
  \small{
    \resizebox{\textwidth}{!}{
    \begin{tabular}{ccccccccccc}
    \toprule
    \multirow{2}[4]{*}{Dataset} & \multirow{2}[4]{*}{Method} & \multirow{2}[4]{*}{Model} & \multicolumn{4}{c}{Validation} & \multicolumn{4}{c}{Test} \\
    % \cmidrule{4-11}
    \cmidrule(lr){4-7} \cmidrule(lr){8-11}
          &       &       & D-MAE $\downarrow$ & D-RMSE $\downarrow$ & C-RMSD $\downarrow$ & E-RMSD $\downarrow$ & D-MAE $\downarrow$ & D-RMSE $\downarrow$ & C-RMSD $\downarrow$ & E-RMSD $\downarrow$ \\
    \midrule
    \multirow{15}[2]{*}{\textbf{Random}} 
          & \multirow{5}[2]{*}{\textbf{3D}} 
          & SE(3)-Transformer \cite{fuchs2020se} & 0.466 & 0.712 & 0.800 & - & 0.467 & 0.774 & 0.802 & - \\
          &       & EGNN  \cite{satorras2021n}  & 0.461 & 0.704 & 0.798 & - & 0.462 & 0.766 & 0.799 & - \\
          &       & ConfOpt-TwoAtom \cite{guan2021energy} & 0.438 & 0.668 & 0.748 & - & 0.438 & 0.670 & 0.749 & - \\
          &       & ConfOpt-ThreeAtom \cite{guan2021energy} & 0.429 & 0.659 & 0.734 & - & 0.430 & 0.661 & 0.736 & - \\
          &       & WGFormer \cite{wang2024wgformer} & \underline{0.391} & \underline{0.649} & \underline{0.662} & - & \underline{0.392} & \underline{0.652} & \underline{0.664} & - \\
    \tightdoublehlineplus
          & \multirow{10}[2]{*}{\textbf{2D}} 
          & RDKit DG \cite{landrum2013rdkit} & 0.581 & 0.930 & 1.054 & 1.864 & 0.582 & 0.932 & 1.055 & 1.872 \\
          &       & RDKit ETKDG \cite{landrum2013rdkit} & 0.575 & 0.941 & 0.998 & 1.700 & 0.576 & 0.942 & 0.999 & 1.710 \\
          &       & GINE \cite{hu2019strategies} & 0.590 & 1.014 & 1.116 & 2.227 & 0.592 & 1.018 & 1.116 & 2.230 \\
          &       & GATv2 \cite{brody2021attentive} & 0.563 & 0.983 & 1.082 & 2.163 & 0.564 & 0.986 & 1.083 & 2.168 \\
          &       & GPS \cite{rampavsek2022recipe}   & 0.528 & 0.909 & 1.036 & 2.089 & 0.529 & 0.911 & 1.038 & 2.094 \\
          &       & GTMGC \cite{xu2023gtmgc} & 0.432 & 0.719 & 0.712 & 1.347 & 0.433 & 0.721 & 0.713 & 1.350 \\
          &       & REBIND \cite{kim2025rebind} & 0.418 & 0.706 & 0.698 & \underline{1.314} & 0.419 & 0.708 & 0.699 & \underline{1.317} \\
          &       & DiffFormer \cite{ye2024differential}   & 0.420 & 0.708 & 0.687 & 1.325 & 0.421 & 0.711 & 0.699 & 1.328 \\
          &       & CFD  \cite{amingcfd}   & 0.397 & 0.682 & 0.684 & - & 0.407 & 0.695 & 0.688 & - \\
          &       & \cellcolor[rgb]{ .851,  .882,  .957}MPSU-Mamba(ours) & 
            \cellcolor[rgb]{ .851,  .882,  .957}\textbf{0.320} & 
            \cellcolor[rgb]{ .851,  .882,  .957}\textbf{0.633} & 
            \cellcolor[rgb]{ .851,  .882,  .957}\textbf{0.593} & 
            \cellcolor[rgb]{ .851,  .882,  .957}\textbf{1.046} & 
            \cellcolor[rgb]{ .851,  .882,  .957}\textbf{0.321} & 
            \cellcolor[rgb]{ .851,  .882,  .957}\textbf{0.636} & 
            \cellcolor[rgb]{ .851,  .882,  .957}\textbf{0.594} & 
            \cellcolor[rgb]{ .851,  .882,  .957}\textbf{1.049} \\
    \midrule \\
    \midrule
    \multirow{14}[2]{*}{\textbf{Scaffold}} 
          & \multirow{5}[2]{*}{\textbf{3D}} 
          & SE(3)-Transformer \cite{fuchs2020se} & 0.460 & 0.676 & 0.775 & - & 0.456 & 0.678 & 0.747 & - \\
          &       & EGNN \cite{satorras2021n}  & 0.448 & 0.666 & 0.758 & - & 0.442 & 0.670 & 0.741 & - \\
          &       & ConfOpt-TwoAtom \cite{guan2021energy} & 0.408 & 0.626 & 0.708 & - & 0.402 & 0.628 & 0.698 & - \\
          &       & ConfOpt-ThreeAtom \cite{guan2021energy} & 0.401 & 0.619 & 0.697 & - & 0.395 & \underline{0.622} & 0.691 & - \\
          &       & WGFormer \cite{wang2024wgformer} & \underline{0.363} & \underline{0.599} & \underline{0.618} & - & \underline{0.360} & \textbf{0.610} & \textbf{0.627} & - \\
    \tightdoublehlineplus
          & \multirow{9}[2]{*}{\textbf{2D}} 
          & RDKit DG \cite{landrum2013rdkit} & 0.542 & 0.872 & 1.001 & 1.751 & 0.524 & 0.857 & 0.973 & 1.780 \\
          &       & RDKit ETKDG \cite{landrum2013rdkit} & 0.531 & 0.874 & 0.928 & 1.565 & 0.511 & 0.859 & 0.898 & 1.595 \\
          &       & GINE \cite{hu2019strategies}  & 0.883 & 1.517 & 1.407 & 3.007 & 1.400 & 2.224 & 1.960 & 4.142 \\
          &       & GATv2 \cite{brody2021attentive} & 0.778 & 1.385 & 1.254 & 2.675 & 1.238 & 2.069 & 1.752 & 3.832 \\
          &       & GPS \cite{rampavsek2022recipe}   & 0.538 & 0.885 & 1.031 & 2.047 & 0.657 & 1.091 & 1.136 & 2.203 \\
          &       & GTMGC \cite{xu2023gtmgc} & 0.406 & 0.675 & 0.678 & 1.318 & 0.400 & 0.679 & 0.693 & 1.275 \\
          &       & REBIND \cite{kim2025rebind} & 0.391 & 0.661 & 0.640 & \underline{1.174} & 0.386 & 0.663 & 0.667 & \textbf{1.182} \\
          &       & DiffFormer \cite{ye2024differential}    & 0.394 & 0.666 & 0.640 & 1.203 & 0.389 & 0.671 & 0.666 & \underline{1.210} \\
          &       & \cellcolor[rgb]{ .851,  .882,  .957}MPSU-Mamba(ours) & 
            \cellcolor[rgb]{ .851,  .882,  .957}\textbf{0.304} & 
            \cellcolor[rgb]{ .851,  .882,  .957}\textbf{0.595} & 
            \cellcolor[rgb]{ .851,  .882,  .957}\textbf{0.574} & 
            \cellcolor[rgb]{ .851,  .882,  .957}\textbf{1.047} & 
            \cellcolor[rgb]{ .851,  .882,  .957}\textbf{0.359} & 
            \cellcolor[rgb]{ .851,  .882,  .957}0.679 & 
            \cellcolor[rgb]{ .851,  .882,  .957}\underline{0.662} & 
            \cellcolor[rgb]{ .851,  .882,  .957}1.218 \\
    \bottomrule
    \end{tabular}%
    }
  }
  \label{table2}
  \vspace{-0.1in}
\end{table*}

To this end, this work explores a Mamba-based framework to obtain a comprehensive understanding of molecular structures. And three specific scanning strategies are devised based on three different perspectives, i.e., atom types, positions, and connections. The performance improvements on the four metrics further demonstrate that a comprehensive understanding of molecular structure is indeed instrumental in accurately predicting conformation. Meanwhile, for small molecules, our proposed scanning strategies could achieve sufficient structure-related information.

\subsection{Prediction on Large-scale Molecular Structures}

\begin{table*}[htbp]
  \centering
  \caption{QM9 generalization performance of cases with few samples.}
  \small{
    \resizebox{\textwidth}{!}{
    \begin{tabular}{cccccccccc}
    \toprule
    \multirow{2}[4]{*}{Split} & \multirow{2}[4]{*}{Method} & \multicolumn{4}{c}{Validation} & \multicolumn{4}{c}{Test} \\
    % \cmidrule{3-10}          
    \cmidrule(lr){3-6} \cmidrule(lr){7-10}
    &       & D-MAE $\downarrow$ & D-RMSE $\downarrow$ & C-RMSD $\downarrow$ & E-RMSD $\downarrow$ & D-MAE $\downarrow$ & D-RMSE $\downarrow$ & C-RMSD $\downarrow$ & E-RMSD $\downarrow$ \\
    \midrule
    \multirow{5}[2]{*}{50\%} 
          & WGFormer & \underline{0.238} & \underline{0.428} & \underline{0.239} & - & \underline{0.241} & \underline{0.434} & \underline{0.248} & - \\
          & GTMGC & 0.316 & 0.509 & 0.471 & 0.917 & 0.318 & 0.512 & 0.475 & 0.926 \\
          & REBIND & 0.246 & 0.444 & 0.306 & 0.567 & 0.248 & 0.448 & 0.311 & 0.582 \\
          & DiffFormer & 0.242 & 0.440 & 0.294 & \underline{0.531} & 0.244 & 0.444 & 0.299 & \underline{0.549} \\
          & \cellcolor[rgb]{ .851,  .882,  .957}MPSU-Mamba(ours) & \cellcolor[rgb]{ .851,  .882,  .957}\textbf{0.092} & \cellcolor[rgb]{ .851,  .882,  .957}\textbf{0.232} & \cellcolor[rgb]{ .851,  .882,  .957}\textbf{0.178} & \cellcolor[rgb]{ .851,  .882,  .957}\textbf{0.350} & \cellcolor[rgb]{ .851,  .882,  .957}\textbf{0.094} & \cellcolor[rgb]{ .851,  .882,  .957}\textbf{0.236} & \cellcolor[rgb]{ .851,  .882,  .957}\textbf{0.181} & \cellcolor[rgb]{ .851,  .882,  .957}\textbf{0.362} \\
    \midrule
    \multirow{5}[2]{*}{25\%} 
          & WGFormer & \underline{0.249} & \underline{0.441} & \underline{0.269} & - & \underline{0.253} & \underline{0.448} & \underline{0.276} & - \\
          & GTMGC & 0.324 & 0.511 & 0.499 & 0.966 & 0.326 & 0.514 & 0.498 & 0.975 \\
          & REBIND & 0.266 & 0.461 & 0.374 & 0.710 & 0.268 & 0.464 & 0.379 & 0.716 \\
          & DiffFormer & 0.254 & 0.448 & 0.329 & \underline{0.626} & 0.257 & 0.452 & 0.333 & \underline{0.640} \\
          & \cellcolor[rgb]{ .851,  .882,  .957}MPSU-Mamba(ours) & \cellcolor[rgb]{ .851,  .882,  .957}\textbf{0.120} & \cellcolor[rgb]{ .851,  .882,  .957}\textbf{0.263} & \cellcolor[rgb]{ .851,  .882,  .957}\textbf{0.227} & \cellcolor[rgb]{ .851,  .882,  .957}\textbf{0.473} & \cellcolor[rgb]{ .851,  .882,  .957}\textbf{0.121} & \cellcolor[rgb]{ .851,  .882,  .957}\textbf{0.267} & \cellcolor[rgb]{ .851,  .882,  .957}\textbf{0.232} & \cellcolor[rgb]{ .851,  .882,  .957}\textbf{0.481} \\
    \midrule
    \multirow{5}[2]{*}{10\%} 
          & WGFormer & \underline{0.287} & \underline{0.504} & \underline{0.396} & - & \underline{0.291} & \underline{0.510} & \underline{0.407} & - \\
          & GTMGC & 0.357 & 0.545 & 0.541 & 1.049 & 0.358 & 0.548 & 0.541 & 1.056 \\
          & REBIND & 0.312 & 0.500 & 0.483 & 0.942 & 0.314 & 0.503 & 0.488 & 0.957 \\
          & DiffFormer & 0.277 & 0.465 & 0.397 & \underline{0.786} & 0.279 & 0.468 & 0.400 & \underline{0.792} \\
          & \cellcolor[rgb]{ .851,  .882,  .957}MPSU-Mamba(ours) & \cellcolor[rgb]{ .851,  .882,  .957}\textbf{0.174} & \cellcolor[rgb]{ .851,  .882,  .957}\textbf{0.325} & \cellcolor[rgb]{ .851,  .882,  .957}\textbf{0.315} & \cellcolor[rgb]{ .851,  .882,  .957}\textbf{0.666} & \cellcolor[rgb]{ .851,  .882,  .957}\textbf{0.174} & \cellcolor[rgb]{ .851,  .882,  .957}\textbf{0.325} & \cellcolor[rgb]{ .851,  .882,  .957}\textbf{0.318} & \cellcolor[rgb]{ .851,  .882,  .957}\textbf{0.675} \\
    \midrule
    \multirow{5}[2]{*}{5\%} 
          & WGFormer & \underline{0.313} & \underline{0.579} & \underline{0.429} & - & \underline{0.319} & \underline{0.588} & \underline{0.440} & - \\
          & GTMGC & 0.392 & 0.573 & 0.585 & 1.135 & 0.393 & 0.574 & 0.583 & 1.138 \\
          & REBIND & 0.347 & 0.538 & 0.576 & 1.119 & 0.350 & 0.544 & 0.586 & 1.140 \\
          & DiffFormer & 0.306 & 0.490 & 0.459 & \underline{0.902} & 0.308 & 0.492 & 0.464 & \underline{0.918} \\
          & \cellcolor[rgb]{ .851,  .882,  .957}MPSU-Mamba(ours) & \cellcolor[rgb]{ .851,  .882,  .957}\textbf{0.230} & \cellcolor[rgb]{ .851,  .882,  .957}\textbf{0.384} & \cellcolor[rgb]{ .851,  .882,  .957}\textbf{0.406} & \cellcolor[rgb]{ .851,  .882,  .957}\textbf{0.833} & \cellcolor[rgb]{ .851,  .882,  .957}\textbf{0.231} & \cellcolor[rgb]{ .851,  .882,  .957}\textbf{0.385} & \cellcolor[rgb]{ .851,  .882,  .957}\textbf{0.405} & \cellcolor[rgb]{ .851,  .882,  .957}\textbf{0.837} \\
    \bottomrule
    \end{tabular}
    }
  }
  \label{table3}
\end{table*}

\begin{table*}[htbp]
  \centering
  \caption{Molecule3D generalization performance of cases with few samples.}
  \small{
    \resizebox{\textwidth}{!}{
    \begin{tabular}{cccccccccc}
    \toprule
    \multirow{2}[4]{*}{Split} & \multirow{2}[4]{*}{Method} & \multicolumn{4}{c}{Validation} & \multicolumn{4}{c}{Test} \\
    \cmidrule(lr){3-6} \cmidrule(lr){7-10}
    &       & D-MAE $\downarrow$ & D-RMSE $\downarrow$ & C-RMSD $\downarrow$ & E-RMSD $\downarrow$ & D-MAE $\downarrow$ & D-RMSE $\downarrow$ & C-RMSD $\downarrow$ & E-RMSD $\downarrow$ \\
    \midrule
    \multirow{5}[2]{*}{25\%} 
          & WGFormer & 0.458 & 0.733 & 0.791 & - & 0.459 & 0.735 & 0.793 & - \\
          & GTMGC & 0.458 & 0.742 & 0.787 & 1.506 & 0.459 & 0.744 & 0.788 & 1.509 \\
          & REBIND & \underline{0.437} & 0.725 & 0.760 & 1.451 & \underline{0.438} & \underline{0.726} & 0.761 & 1.454 \\
          & DiffFormer & 0.438 & \underline{0.724} & \underline{0.737} & \underline{1.430} & \underline{0.438} & \underline{0.726} & \underline{0.738} & \underline{1.433} \\
          & \cellcolor[rgb]{ .851,  .882,  .957}MPSU-Mamba(ours) & 
          \cellcolor[rgb]{ .851,  .882,  .957}\textbf{0.370} & 
          \cellcolor[rgb]{ .851,  .882,  .957}\textbf{0.689} & 
          \cellcolor[rgb]{ .851,  .882,  .957}\textbf{0.704} & 
          \cellcolor[rgb]{ .851,  .882,  .957}\textbf{1.300} & 
          \cellcolor[rgb]{ .851,  .882,  .957}\textbf{0.372} & 
          \cellcolor[rgb]{ .851,  .882,  .957}\textbf{0.691} & 
          \cellcolor[rgb]{ .851,  .882,  .957}\textbf{0.705} & 
          \cellcolor[rgb]{ .851,  .882,  .957}\textbf{1.303} \\
    \midrule
    \multirow{5}[2]{*}{10\%} 
          & WGFormer & 0.468 & 0.745 & 0.815 & - & 0.469 & 0.747 & 0.816 & - \\
          & GTMGC & 0.479 & 0.760 & 0.828 & 1.574 & 0.480 & 0.762 & 0.829 & 1.577 \\
          & REBIND & 0.453 & 0.742 & 0.789 & \underline{1.510} & 0.454 & 0.744 & 0.790 & \underline{1.513} \\
          & DiffFormer & \underline{0.451} & \underline{0.733} & \underline{0.780} & 1.512 & \underline{0.451} & \underline{0.735} & \underline{0.782} & 1.515 \\
          & \cellcolor[rgb]{ .851,  .882,  .957}MPSU-Mamba(ours) & 
          \cellcolor[rgb]{ .851,  .882,  .957}\textbf{0.411} & 
          \cellcolor[rgb]{ .851,  .882,  .957}\textbf{0.731} & 
          \cellcolor[rgb]{ .851,  .882,  .957}\textbf{0.761} & 
          \cellcolor[rgb]{ .851,  .882,  .957}\textbf{1.441} & 
          \cellcolor[rgb]{ .851,  .882,  .957}\textbf{0.412} & 
          \cellcolor[rgb]{ .851,  .882,  .957}\textbf{0.733} & 
          \cellcolor[rgb]{ .851,  .882,  .957}\textbf{0.762} & 
          \cellcolor[rgb]{ .851,  .882,  .957}\textbf{1.445} \\
    \midrule
    \multirow{5}[2]{*}{5\%} 
          & WGFormer & 0.476 & 0.758 & 0.828 & - & 0.477 & 0.760 & 0.829 & - \\
          & GTMGC & 0.494 & 0.773 & 0.863 & 1.629 & 0.495 & 0.775 & 0.864 & 1.632 \\
          & REBIND & 0.465 & 0.752 & 0.839 & 1.613 & 0.466 & 0.754 & 0.840 & 1.616 \\
          & DiffFormer & \underline{0.464} & \underline{0.743} & \underline{0.808} & \underline{1.576} & \underline{0.464} & \underline{0.745} & \underline{0.809} & \underline{1.579} \\
          & \cellcolor[rgb]{ .851,  .882,  .957}MPSU-Mamba(ours) & 
          \cellcolor[rgb]{ .851,  .882,  .957}\textbf{0.449} & 
          \cellcolor[rgb]{ .851,  .882,  .957}\textbf{0.742} & 
          \cellcolor[rgb]{ .851,  .882,  .957}\textbf{0.825} & 
          \cellcolor[rgb]{ .851,  .882,  .957}\textbf{1.544} & 
          \cellcolor[rgb]{ .851,  .882,  .957}\textbf{0.450} & 
          \cellcolor[rgb]{ .851,  .882,  .957}\textbf{0.745} & 
          \cellcolor[rgb]{ .851,  .882,  .957}\textbf{0.826} & 
          \cellcolor[rgb]{ .851,  .882,  .957}\textbf{1.546} \\
    \bottomrule
    \end{tabular}
    }
  }
  \label{table4}
\end{table*}

Next, we further verify the prediction results of molecular conformation on large-scale datasets. Here, we utilize Molecule3D dataset \cite{xu2021molecule3d}. And this dataset comprises approximately 4 million molecules, each with its own 2D molecular graph, ground-state 3D geometric structure, and four additional quantum properties. Meanwhile, it utilizes two splitting strategies: random and scaffold splittings.

Table \ref{table2} shows the prediction results. Here, the model architecture and hyper-parameter settings are the same as those of the model trained on QM9 dataset. For the Random split, based on the given four metrics, our method still obtains the best prediction performance. Particularly, compared with the state-of-the-art 2D method, i.e., REBIND \cite{kim2025rebind}, our method separately outperforms it by around 9.8\%, 7.2\%, 10.5\%, and 26.8\%. Meanwhile, compared with 3D methods, the proposed method also achieves the best results. For example, our method is around 7.1\% higher than WGFormer \cite{wang2024wgformer}. These improvements further indicate that comprehensively understanding molecular structures could significantly strengthen the performance of conformation prediction. Moreover, this also demonstrates that designing proper scanning mechanisms is instrumental in capturing plentiful structure-relevant characteristics. For the Scaffold split, we observe that for the validation and test sets, based on 2D mechanisms, our method almost obtains the best performance on the four metrics. Similarly, on the validation set, our method outperforms REBIND \cite{kim2025rebind} by around 8.7\%, 6.7\%, 6.6\%, and 12.7\%. This further shows that for the 2D-to-3D conformation prediction, only stacking multi-layer attention may neglect certain critical information about the overall molecular structure, weakening the prediction performance. Besides, we can also observe that for D-RMSE and C-RMSD metrics, our performance is weaker than WGFormer \cite{wang2024wgformer}. However, WGFormer \cite{wang2024wgformer} first utilizes RDKit to obtain the initial 3D conformation for the following processes. Here, the input is converted from 2D to 3D, which may increase the complexity of optimization. Therefore, existing experimental results could demonstrate that utilizing Mamba to perform multi-perspective understanding is beneficial for capturing sufficient molecular structure-relevant information.

\subsection{Generalization on cases with Few Training Samples}

In practice, for certain specific applications, it is difficult to collect large-scale data with diverse molecular structures and corresponding annotations for training. Therefore, it is necessary to verify the generalization performance on small sample training datasets. Here, based on QM9 and Molecule3D datasets, we randomly select 25\%, 10\%, and 5\% of the overall training data. And the validation and test data are kept unchanged. Four state-of-the-art methods, i.e., GTMGC \cite{xu2023gtmgc}, WGFormer \cite{wang2024wgformer}, REBIND \cite{kim2025rebind}, and DiffFormer \cite{ye2024differential}, are used for comparison. Table \ref{table3} shows the QM9 results. We can observe that as the amount of training data decreases, the performance of the model gradually deteriorates.Taking the D-MAE metric as an example, when the data amount varies from 50\% to 5\%, the result decreases from 9.2\% to 23.0\%, which indicates the necessity of evaluation on cases with few training samples. Meanwhile, we can see that for all splits, our method still obtains the best performance of conformation prediction. Particularly, when the data amount is 5\%, our method achieves the D-MAE prediction performance that is similar to REBIND \cite{kim2025rebind} trained based on the overall data. This further shows that understanding the structure information is efficient in cases with few training samples. Besides, DiffFormer \cite{ye2024differential} is a newly proposed Transformer architecture, aiming to amplify attention to the relevant context while canceling noise. Specifically, the differential attention mechanism calculates attention scores as the difference between two separate softmax attention maps. We consider that enlarging attention to the relevant context could capture more structure-aware information. Experimental results also demonstrate that calculating the differential attention map could improve the performance of conformation prediction. Whereas, our method still outperforms DiffFormer. The reason lies in that for our method, during iteration, each state could possess longer context information. Meanwhile, each dedicated scanning perspective represents a different understanding level of molecular structures. By fusing multiple perspectives, the critical structural information could be captured effectively, enhancing the prediction ability of molecular conformation.

In Table \ref{table4}, we further evaluate the generalization on Molecule3D dataset. Since the Molecule3D is a large-scale dataset, we focus on the result of using 5\%, 10\%,  and 25\% training data. We can also observe that compared with state-of-the-art methods, our approach still achieves superior performance. For example, the D-MAE result of our method is around 4.2\% higher than REBIND \cite{kim2025rebind} and 3.9\% higher than DiffFormer \cite{ye2024differential}. This could demonstrate the effectiveness of our method. However, we observe that the error is still high, e.g., D-MAE is higher than 40\%. This indicates that predicting molecular conformation with the lowest energy is filled with challenges, which requires continuous exploration.

\subsection{Analysis of Visual Comparison}

\begin{figure*}[p]
  \centering
  \includegraphics[width=\textwidth]{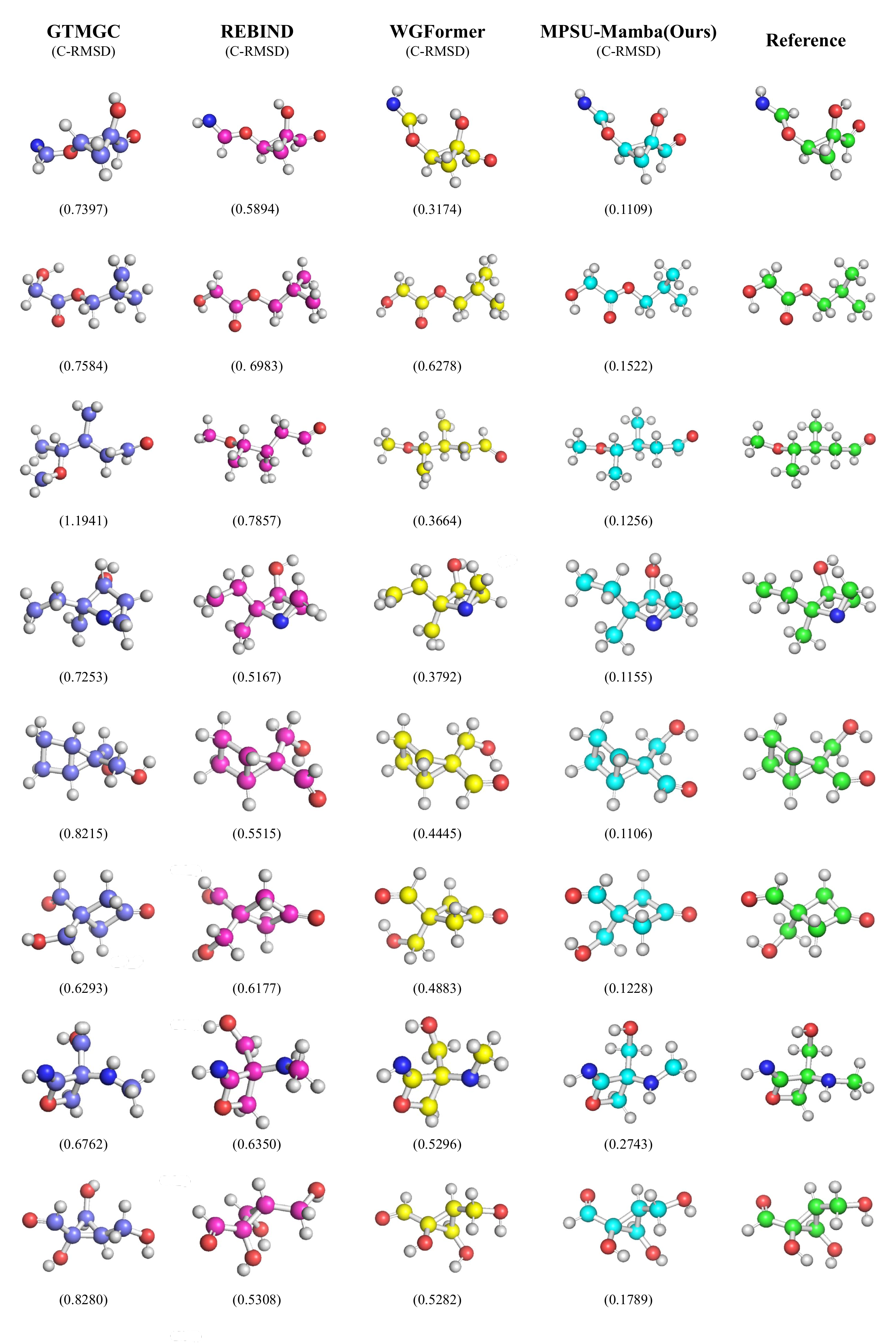}
  \vspace{-0.3in}
  \caption{Visual comparison between our MPSU-Mamba and existing state-of-the-art methods.}
  \label{fig4}
\end{figure*}

\begin{figure*}[p]
  \centering
  \includegraphics[width=\textwidth]{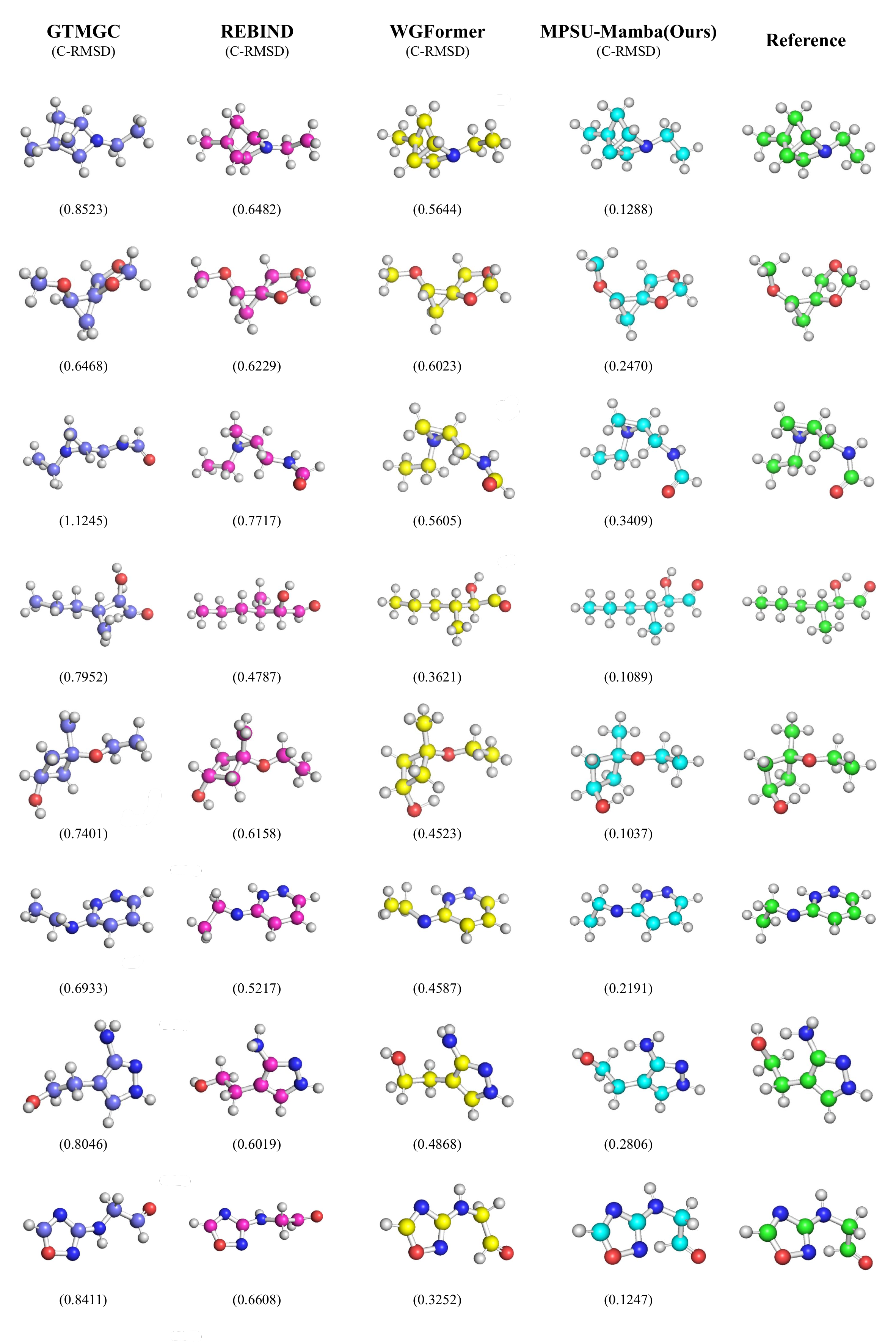}
  \vspace{-0.3in}
  \caption{More comparison results between our MPSU-Mamba and existing state-of-the-art methods.}
  \label{fig5}
\end{figure*}

In Fig. \ref{fig4} and \ref{fig5}, based on QM9 dataset, we show some compared visual results between our method and other state-of-the-art methods, i.e., GTMGC \cite{xu2023gtmgc}, REBIND \cite{kim2025rebind}, and WGFormer \cite{wang2024wgformer}. And the corresponding C-RMSD values are provided under the predicted results. Firstly, we can directly see that due to differences in atom types, positions, and connections between atoms, molecular structures exhibit diverse and complex characteristics, posing significant challenges for the accurate prediction of molecular conformation.  Particularly, the presence of ring structures further increases the complexity, resulting in inaccurate predictions. This indicates that a comprehensive understanding of molecular structures is important for improving the performance of downstream tasks. Next, in Fig. \ref{fig4} and \ref{fig5}, we observe that compared with GTMGC, REBIND obtains more precise predictions. This shows that designing proper attention mechanisms for capturing critical atoms and relationships is instrumental in obtaining plentiful conformation-related information. Meanwhile, we observe that compared with GTMGC and REBIND, WGFormer achieves better results, demonstrating that by estimating the initial 3D positions, transferring the 2D-to-3D conformation prediction to the 3D-to-3D problem is effective and is worth further exploration. Finally, we can observe that compared with GTMGC, REBIND, and WGFormer, for molecules with diverse and complex structures, our method obtains the smallest C-RMSD scores and the most accurate predictions. Taking the fifth prediction result as an example, we can see that this molecule contains a ring and a complex structure. And predictions from other methods all have some errors, e.g., there exist errors in the folding of the ring. Whereas, our method could effectively capture critical structure information and produce precise 3D molecular conformation. This still demonstrates that a comprehensive understanding of molecular structures is conducive to obtaining sufficient conformation-aware information. Meanwhile, this further indicates that by capturing long-range dependencies, our Mamba-based mechanism could build a deep perception of molecular structures.

\begin{figure*}[h]
  \centering
  \includegraphics[width=\textwidth]{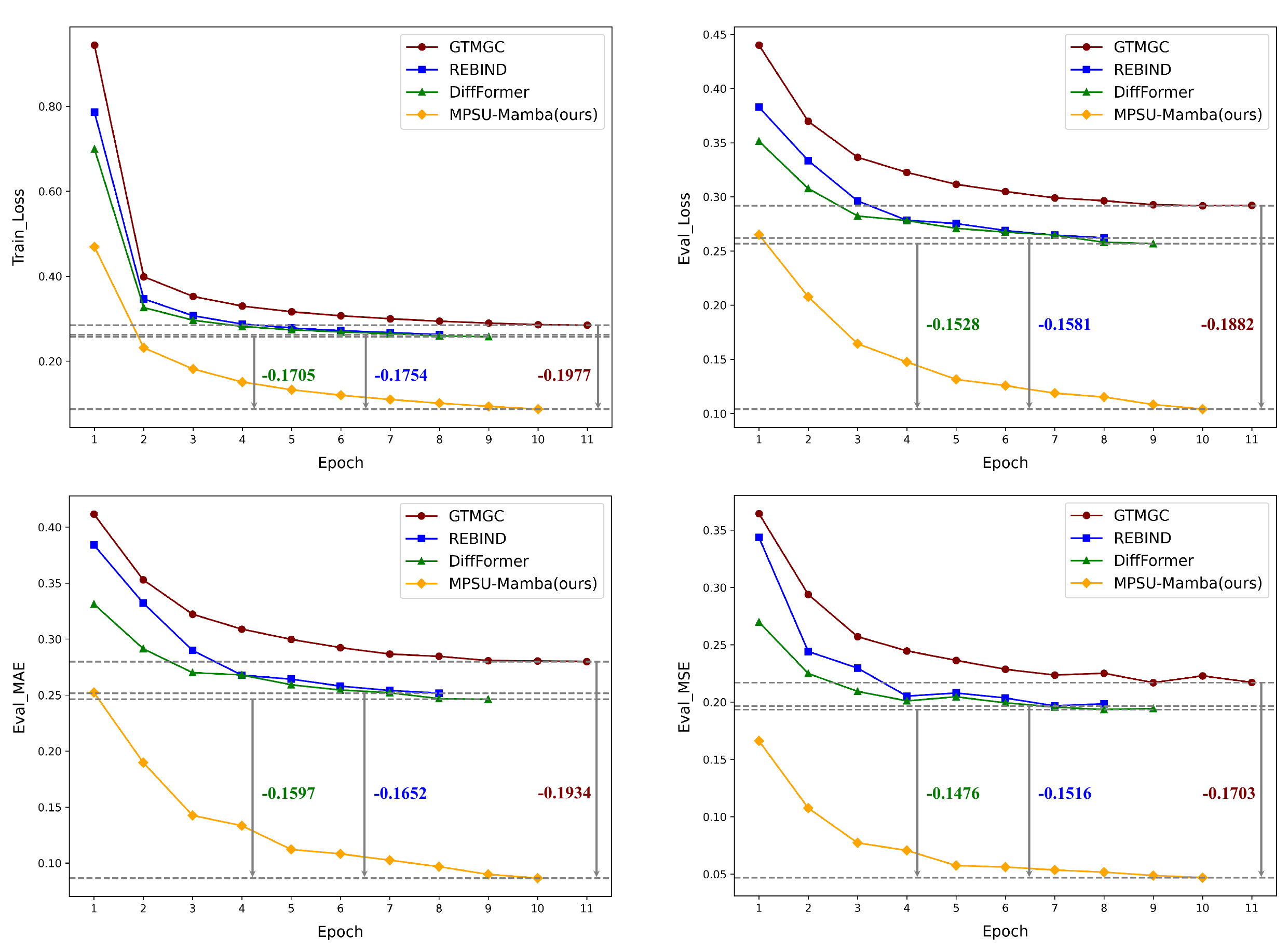}
  \vspace{-0.2in}
  \caption{Training details based on QM9 dataset. For 2D methods, the loss curve of our method is significant lower than GTMGC, RIBIND, and DiffFormer.}
  \label{fig6}
\end{figure*}

\begin{figure*}[h]
  \centering
  \includegraphics[width=\textwidth]{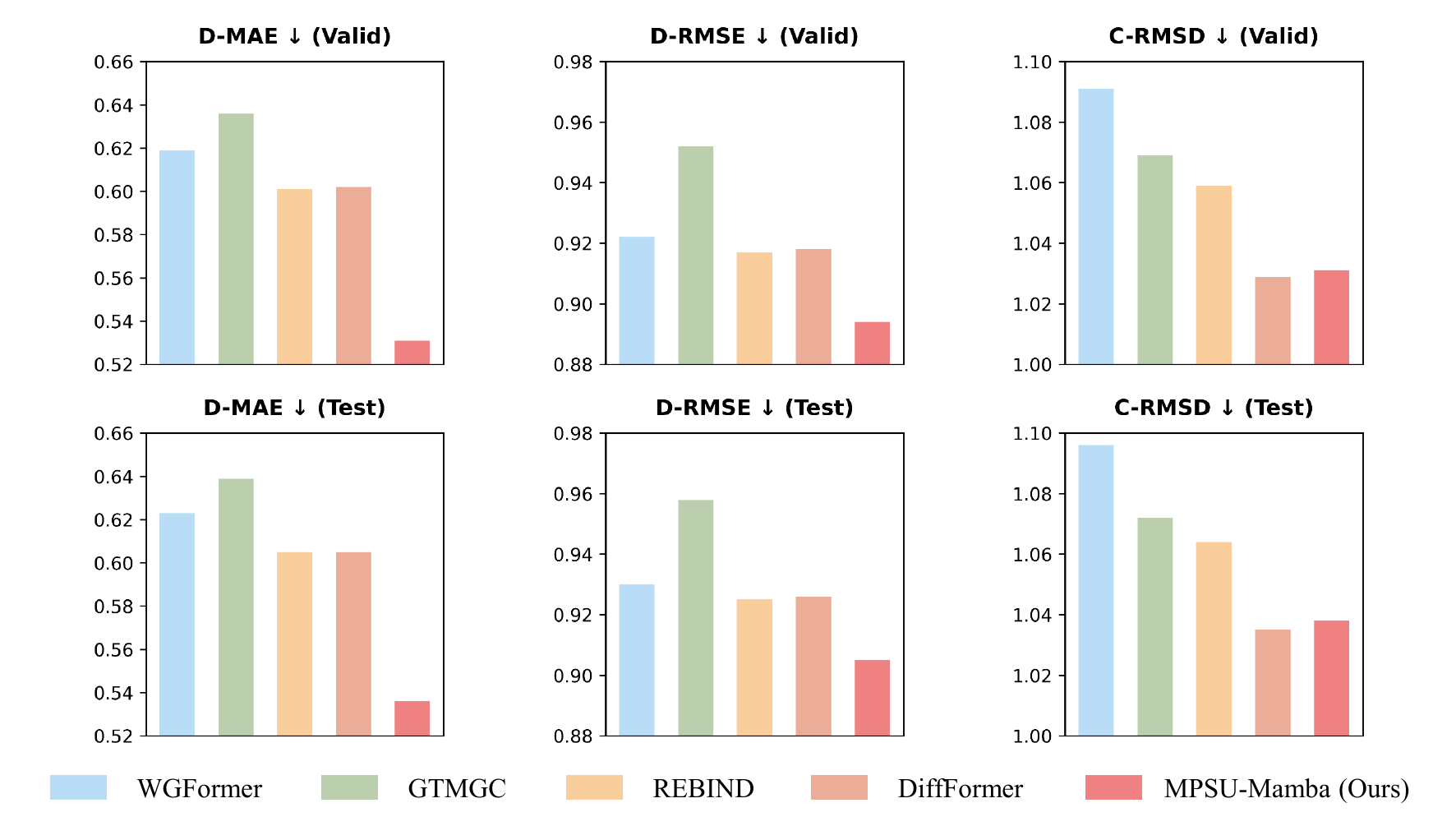}
  \vspace{-0.2in}
  \caption{In general, molecules with more atoms possess complex structures. Conformation prediction also becomes more difficult. Thus, we analyze the performance with more than 40 atoms.}
  \label{fig7}
\end{figure*}

\begin{figure*}[h]
  \centering
  \includegraphics[width=\textwidth]{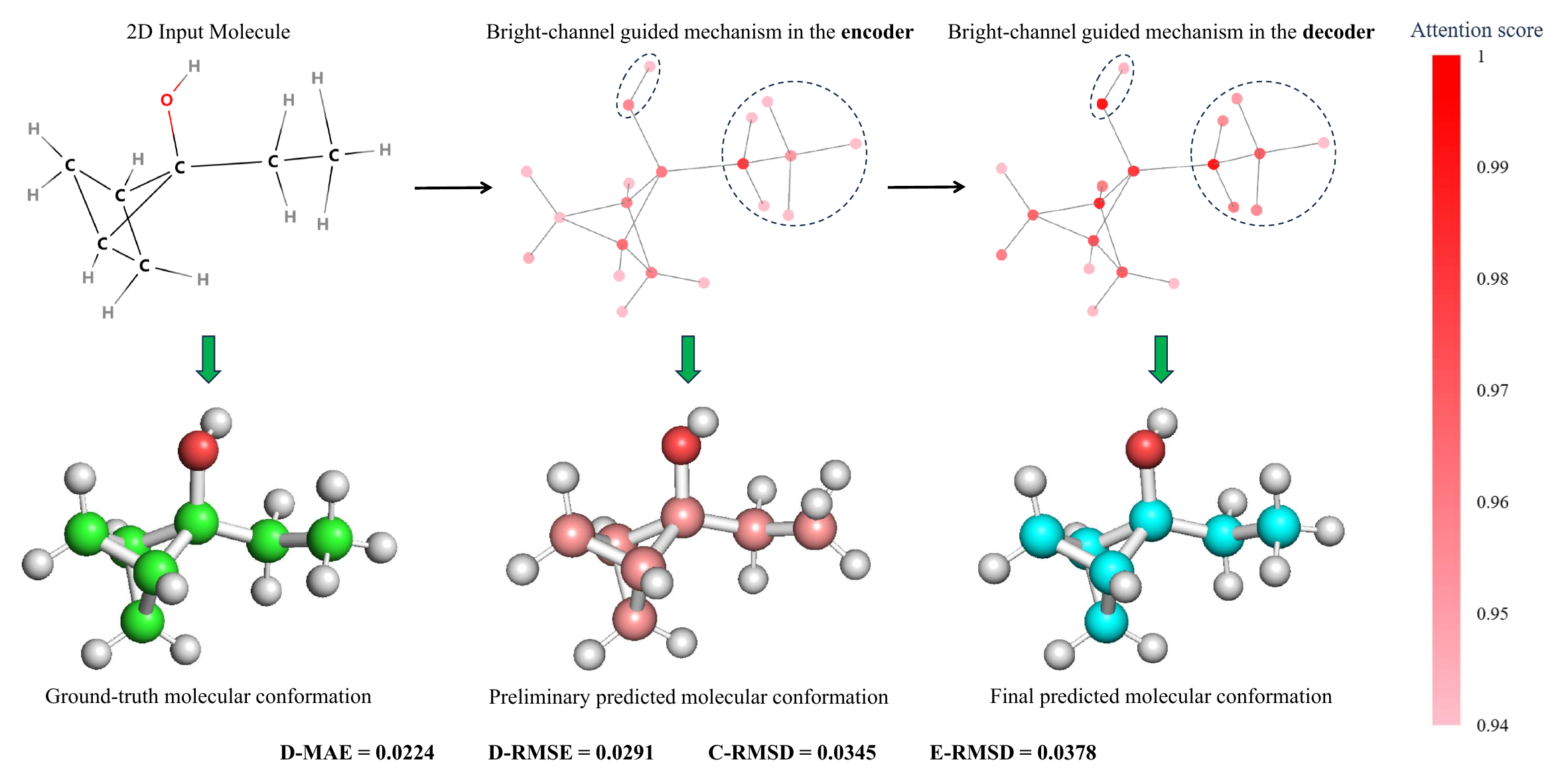}
  \vspace{-0.2in}
  \caption{Visualization example of bright-channel guided mechanism. Here, we separately select the guided layer in the encoder and decoder for visualization.}
  \label{fig8}
\end{figure*}

\subsection{More Analysis of MPSU-Mamba}

Compared with existing attention-based methods, our method is a different architecture that employs Mamba-based mechanism. Thus, it is neccessary to analyze the optimization details. In Fig. \ref{fig6}, we provide corresponding optimization curves. Here, we show loss curves calculated based on the training and validation data. Meanwhile, we also show MAE and MSE values computed based on validation data. We can observe that compared with GTMGC, REBIND, and DiffFormer, the optimization speed of our Mamba-based method is significantly faster than these attention-based methods. This also demonstrates that the computation complexity of Mamba is indeed lower than attention. Meanwhile, we can see that during training, the MAE and MSE values of our method decreases significantly. For example, compared with GTMGC, REBIND, and DiffFormer, the MAE value of our method separately decreases 0.1934, 0.1652, and 0.1587. And the MSE value of our method respectively decreases 0.1703, 0.1516, and 0.1476, further indicating the effectiveness of our method.

\begin{figure*}[h]
  \centering
  \includegraphics[width=\textwidth]{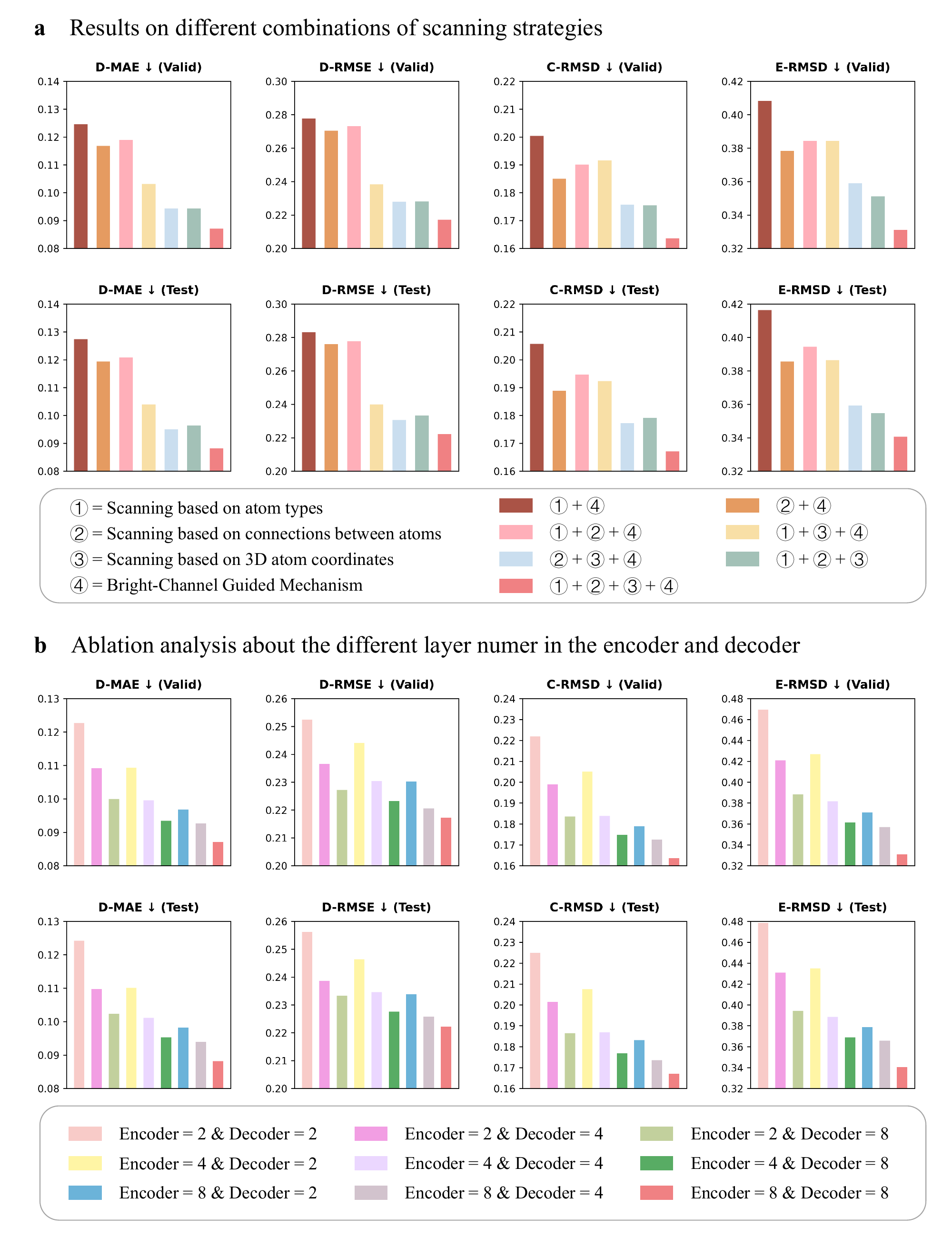}
  \vspace{-0.3in}
  \caption{Based on the validation and test data, we separately perform ablation studies of scanning strategies, encoder layer number, and decoder layer number. Here, we only change these hyper-parameters. Other settings including optimization details are kept unchanged.}
  \label{fig9}
\end{figure*}

In general, molecular structures are related with the atom number. Molecules with more atoms usually involve complex structures, which poses severe challenges for conformation prediction. To this end, we further analyze the performance with more than 40 atoms. In Fig. \ref{fig7}, we show the comparison results. We can see that based on D-MAE, D-RMSE, and C-RMSD metrics, performances of all state-of-the-art methods decrease significantly. This indicates that for molecules with more atoms, conformation prediction is filled with difficulties. However, we can observe that our method still achieve superior performance. Particularly, for D-MAE and D-RMSE metrics, our method obtains the best performance. This could demonstrate that our method is instrumental in constructing a complete understanding of molecular structures, promoting producing accurate conformation predictions.

In Fig. \ref{fig1}, to focus on critical conformation-related atoms, we exploit a dedicated bright-channel guided mechanism. Here, we make a visualization analysis. Particularly, in the encoder and decoder, we separately stack multiple guided layers. In Fig. \ref{fig8}, we can observe that in the encoder, the bright-channel guided mechanism mainly focuses on atoms in the backbone. And in the decoder, the bright-channel guided mechanism pays attention to atoms in the side branches, e.g., atoms located in the circles. Particularly, we can see that after the encoding process, the 3D position of atoms in the circles are gradually adjusted towards to the correct direction. Thus, combining the focused results in the encoder and decoder, our method could achieve accurate conformation predictions. This demonstrates that by using significant elements in the representations for computing corresponding weights, it is beneficial for capturing critical structure-related information, improving the prediction accuracy.

For our method, designing efficient scanning strategies and stacking proper layers in the encoder and decoder are important to build a comprehensive understanding of molecular structures. In Fig. \ref{fig9}, we provide corresponding ablation results. Particularly, in Fig. \ref{fig9}a, we can directly observe that only using one-type scaning could not build a complete perception of molecular structures, which reduces the prediction performance. Besides, combining these three scanning mechanisms is beneficial for constructing a comprehensive molecular understanding, achieving superior results. Finally, in Fig. \ref{fig9}b, we further analyze the impact of stacking different Mamba layers in the encoder and decoder. We can see that stacking different layers affect the performance. For our method, when the layer numer in the encoder and decoder is set to 8, the corresponding performance of conformation prediction is the best.

\section{Discussion}

Predicting the lowest energy state of a molecule accurately, i.e., the molecular ground-state conformation, is meaningful for biochemical research. This work demonstrates that a comprehensive understanding of molecular structure is instrumental in effectively capturing ground-state conformation involving critical property information. In this paper, by constructing proper scanning strategies, a Mamba-driven approach could establish complete structure information, boosting the prediction performance of conformation significantly. Extensive experimental results on multiple benchmarks indicate the superiorities of this method.

Since molecular structures generally exhibit diverse and complex characteristics, achieving precise structure perception is full of challenges. To this end, different attention mechanisms are explored to gradually focus on critical atoms. However, based on the attention weights, it is difficult to establish a complete structure information, which weakens the prediction performance. Motivated by the powerful long-sequence modeling capabilities of Mamba and its great efficiency, we explore employing Mamba to construct a comprehensive perception of molecular structure.

However, in order to effectively leverage Mamba, it is critical to define proper scanning strategies. In the experiments, we observe that using unproper scanning strategies could not improve the prediction performance. In fact, certain scanning mechanisms can significantly degrade the result of conformation prediction. This indicates that using a simple and random scanning mechanism could not capture complete structure content. After numerous attempts, we strive to design some dedicated scanning strategies based on components of molecular structures. In general, molecular structures could be directly considered to consist of three elements, i.e., atom types, atom positions, and connections between atoms. Therefore, considering the three elements, we design three different scanning strategies and propose MPSU-Mamba. Extensive experimental results on two datasets with different settings indicate that using these three scanning strategies could strengthen the structural perception.

Future work continues paying attention to exploring more scanning strategies and scaling the framework to understand biomacromolecules. Beyond conformation prediction, our method could be used for molecule or protein design. In general, for molecule design, it is important to make structures of designed molecules fulfill specific functions. Therefore, using our method could effectively promote understanding molecular structures and improve the effectiveness of designed molecules.

\section{Method}

As shown in Fig.~\ref{fig1}, given an 2D input molecule, we first flatten it randomly into a sequence of atoms and then extract the corresponding embedding features through a linear projector. 
These embedding features are then combined with Laplacian Positional Encoding \cite{kim2025rebind} to construct the preliminary node embedding representation \(\mathbf{X} \in \mathbb{R}^{\rm N \times D}\), where ${\rm N}$ and ${\rm D}$ separately indicate the atom number and feature dimension. Here, MPSU-Mamba follows an encoder-decoder architecture. 
In the encoder, we first utilize two scanning strategies based on atom types and connections between atoms, through which molecular representations are extracted via multiple MPSU-Mamba layers. 
Next, based on the encoded representation, a conformation prediction head is defined to perform an preliminary prediction of 3D atom positions. 
To account for atom positions, in the decoder, we additionally employ different scanning orders based on X, Y, and Z coordinates, while maintaining the original scanning strategies.
Furthermore, multiple bright-channel guided operations are utilized to gradually focus on critical conformation-related atoms. 
Finally, we exploit another conformation prediction head to achieve the final 3D molecular conformation $\mathbf{G} \in \mathbb{R}^{\rm N \times 3}$. 
% The detailed workflow of our method is provided in Algorithm~\ref {algorithm4}.

\subsection{Preliminaries}

The State Space Model (SSM) is commonly used to describe the dynamics of continuous systems, i.e., transforming a 1D continuous input sequence \(x(t) \in \mathbb{R}\) into an output sequence \(y(t) \in \mathbb{R}\) through a learnable latent state \(\mathbf{h}(t) \in \mathbb{R}^Q\). The mathematical formulation of the 
SSM is as follows:
\begin{equation}
\mathbf{h}'(t) = \mathbf{A}\mathbf{h}(t) + \mathbf{B}x(t), 
\quad 
y(t) = \mathbf{C}\mathbf{h}(t) + \mathbf{D}x(t),
\label{eq1}
\end{equation}
where \(\mathbf{A} \in \mathbb{R}^{Q \times Q}\), \(\mathbf{B} \in \mathbb{R}^{Q \times 1}\), \(\mathbf{C} \in \mathbb{R}^{1 \times Q}\), and \(\mathbf{D} \in \mathbb{R}\) are learnable parameters.

To further handle discrete input sequences, the continuous parameters \(\mathbf{A}\), \(\mathbf{B}\), and \(\mathbf{C}\) in Eq.~\ref{eq1} could be converted into discrete parameters. The parameter \(\mathbf{D}\) is omitted, as it can be regarded as a skip connection. Assuming the timescale parameter is \(\Delta \in \mathbb{R}\), applying the zero-order hold rule yields the discrete parameters:
\begin{equation}
\bar{\mathbf{A}} = \exp(\Delta \mathbf{A}), \quad
\bar{\mathbf{B}} = (\Delta \mathbf{A})^{-1} \cdot \big(\exp(\Delta \mathbf{A}) - I\big) \cdot (\Delta \mathbf{B}), \quad
\bar{\mathbf{C}} = \mathbf{C},
\label{eq2}
\end{equation}
where \(I\) denotes the identity matrix. 
% The Eq.~\ref{eq1} can then be expressed with discrete parameters as:
Thus, Eq.~\eqref{eq1} can be re-represented as follows:
\begin{equation}
\mathbf{h}(t) = \bar{\mathbf{A}}\mathbf{h}(t - 1) + \bar{\mathbf{B}}x(t), \quad
y(t) = \bar{\mathbf{C}}\mathbf{h}(t).
\label{eq3}
\end{equation}

To overcome the linear time-invariant property of SSMs, Mamba~\cite{gu2023mamba} introduces a selective scanning mechanism, allowing input-dependent sequence processing.
Specifically, the model parameters \(\mathbf{B}\), \(\mathbf{C}\), and \(\Delta\) are the functions of the input  \(x(t)\), allowing them to dynamically adjust according to the input. This enables the model to filter out irrelevant information while retaining important features, thereby capturing plentiful complex input dependencies.
Since Eq.~\eqref{eq3} is difficult to compute parallelly, Mamba calculates the entire sequence’s output simultaneously through a global convolution:
% \vspace{-0.1in}
\begin{equation}
\mathbf{y} = \bar{\mathbf{K}} \ast \mathbf{x}, 
\quad
\bar{\mathbf{K}} \in \mathbb{R}^{M} 
= (\overline{\mathbf{C}\mathbf{B}},\,
\overline{\mathbf{C}\mathbf{A}\mathbf{B}},\,
\ldots,\,
\overline{\mathbf{C}\mathbf{A}}^{M-1}\overline{\mathbf{B}}),
\label{eq4}
\end{equation}
% \vspace{-0.1in}
where \(\bar{\mathbf{K}}\) is the kernel of the global convolution. \(\mathbf{x} \) and \(\mathbf{y} \) separately denote the input and output sequences. $M$ is the length of the input sequence, and \(\ast\) indicates the convolution operation.

A comprehensive understanding of molecular structures is important for accurate conformation predictions. Therefore, as shown in Fig. \ref{fig1}, this work first explores designing dedicated scanning strategies and proposes a MPSU-Mamba method that follows the encoder-decoder architecture.

\subsection{MPSU-Mamba Encoder for Preliminary Prediction}

As shown in Fig. \ref{fig1}a and b, based on the preliminary node embedding representation $\mathbf{X} \in \mathbb{R}^{\rm N \times D}$, we first utilize a bright-channel guided mechanism to obtain ${\rm X}_{in} \in \mathbb{R}^{\rm N \times D}$. Next, \textbf{L} MPSU-Mamba operations are used to extract the encoding representation. Concretely, in Fig. \ref{fig1}c, taking ${\rm X}_{in}^{i} \in \mathbb{R}^{\rm N \times D}$ as the input (where $i=1, 2, 3, \cdots, \mathbf{L}$ and ${\rm X}_{in}^1 = {\rm X}_{in}$), two different linear projectors are exploited to calculate ${\rm x}_{i} \in \mathbb{R}^{\rm N\times D}$ and ${\rm z}_{i} \in \mathbb{R}^{\rm N\times D}$. Then, a Depthwise 1D convolution followed by a SiLU activation is applied to ${\rm x}_i$ to obtain ${\rm x}_i' \in \mathbb{R}^{\rm N\times D}$. To build an initial understanding of molecular structures, two dedicated scanning strategies are selected based on atom types and connections between atoms. Meanwhile, we process ${\rm x}_i'$ from the forward direction:
\begin{equation}
\mathbf{h}_{i}^{o}(t) = \mathbf{\bar{A}}_{i}\mathbf{h}^{o}_{i}(t - 1) + \mathbf{\bar{B}}_{i}{\rm x}_i', \quad
\mathbf{y}_{i}^{o} = \mathbf{\bar{C}}_{i}\mathbf{h}^{o}_{i}(t),
\label{eq5}
\end{equation}
where $t=1,2,3, \cdots, {\rm N}$. $\mathbf{\bar{A}}_{i}$, $\mathbf{\bar{B}}_{i}$, and $\mathbf{\bar{C}}_{i}$ represents the learnable parameters for the $i$-th MPSU-Mamba operation. As shown in Fig. \ref{fig1}a, based on atom type and connections, three different scanning orders are produced. For $o$-th order, the MPSU-Mamba processes are the same. Here, it is worth noting that during encoding, the forward direction is only used. Finally, we can obtain three outputs, i.e., $\mathbf{y}_{i}^{1} \in \mathbb{R}^{\rm N \times D}$, $\mathbf{y}_{i}^{2} \in \mathbb{R}^{\rm N \times D}$, and $\mathbf{y}_{i}^{3} \in \mathbb{R}^{\rm N \times D}$. And the concatenation result of $\mathbf{y}_{i}^{1}$, $\mathbf{y}_{i}^{2}$, and $\mathbf{y}_{i}^{3}$ is input into a linear projection and a LayerNorm operation, which outputs $\mathbf{y}_{i} \in \mathbb{R}^{\rm N\times D}$. Obviously, $\mathbf{y}_{i}$ has involved the information about atom types and connections. Particularly, it also considers the impact of the ring structure.

Next, in Fig. \ref{fig1}c, a fusion operation is performed on $\mathbf{y}_{i}$ and the activation result of ${\rm z}_{i}$. By means of a linear project, we can obtain the output ${\rm X}_{out}^{i} \in \mathbb{R}^{\rm N \times D}$ of the current operation. Through multiple MPSU-Mamba operations, the encoder output ${\rm X}_{out}^{\bf{L}} \in \mathbb{R}^{\rm N \times D}$ is taken as the input of the bright-channel guided mechanism, obtaining $X_{e} \in \mathbb{R}^{\rm N \times D}$. Finally, through a prediction head, the preliminary result $G_{e} \in \mathbb{R}^{\rm N \times 3}$ is used to calculate the corresponding loss $\mathcal{L}_{encoder}$. The processes are shown as follows:
\begin{equation}
\mathcal{L}_{encoder}
= \frac{\displaystyle\sum_{p=1}^{\rm N}\sum_{q=1}^{3} 
\left\vert G_e[p,q] - \hat{\mathbf{G}}[p,q] \right\vert}{\rm N \times 3},
\label{eq6}
\end{equation}
where $\hat{\mathbf{G}} \in \mathbb{R}^{\rm N \times 3}$ represents the ground-truth molecular conformation. Through these processes, the MPSU-Mamba encoder is promoted to gradually construct a preliminary perception of molecular structures, boosting the prediction accuracy of the decoder.

\subsection{Bright-Channel Guided Mechanism}

In general, conformation refers to the specific shape of a molecule, reflecting the actual relative positions of the atoms. Each atom contributes differently to the final conformation. Therefore, to further improve the prediction accuracy, we design a dedicated bright-channel guided mechanism.

Specifically, as shown in Fig. \ref{fig1}b, taking $\mathbf{X} \in \mathbb{R}^{\rm N \times D}$ as the input, we first compute the maximum activation along the channel dimension. By means of a Sigmoid operation, a sequence of atom-level weights can be obtained as follows: 
\begin{equation}
X_{in} = \mathbf{X} \odot \sigma(\text{max}(\textbf{X},\, \text{dim}=-1)),
\label{eq7}
\end{equation}
where \(\sigma(\cdot)\) denotes the sigmoid function, and \(\odot\) indicates element-wise multiplication. $X_{in} \in \mathbb{R}^{\rm N \times D}$ is the corresponding guided representation. We assume that in a high-dimensional representation space, the maximized channel may involve significant task-relevant information. Therefore, through optimization, this operation is instrumental in gradually focusing on critial conformation-aware atoms. As shown in Fig. \ref{fig8} and \ref{fig9}a, the visualization and ablation results demonstrate that the bright-channel guide operation could strengthen the accuracy of conformation prediction.

\subsection{MPSU-Mamba Decoder for Final Prediction}

As shown in Fig. \ref{fig1}a, the decoder processes are similar to the encoder. Concretely, $X_{e}$ is still combined with Laplacian Position Encoding to achieve $\mathbf{I} \in \mathbb{R}^{\rm N \times D}$. Through a bright-channel guided operation, we can obtain the input $I_{in} \in \mathbb{R}^{\rm N \times D}$ of the decoder network. Next, $\mathbf{L}$ MPSU-Mamba operations are used to extract the decoding representation. The processes are the same as the encoder. 

After the encoding process, we can obtain the preliminary 3D positions $G_{e}$. Since conformation reflects the actual relative positions of the atoms, it is meaningful to consider the spatial information, which is beneficial for enhancing the understanding of molecular structures. To this end, besides using type- and connection-based scannings, during decoding, we additionally employ the scanning strategy based on X, Y, and Z coordinates from the forward and backward directions, resulting in 9 different scanning orders (as shown in Fig. \ref{fig1}a). For each scanning order, the MPSU-Mamba processes are the same. It is worth noting that for our method, we only employ the bidirectional scanning for the position information. Finally, for each MPSU-Mamba operation, nine outputs with the same size can be obtained. Then, the concatenation result of the nine results is still input into a linear projection and a LayerNorm operation, which outputs $\mathcal{O}_{i} \in \mathbb{R}^{\rm N \times D}$. Obviously, $\mathcal{O}_{i}$ has involved type, connection, and position-based information, constructing a relatively complete structure information. 

Next, the same fusion operation as the encoder is performed on $\mathcal{O}_{i}$ and the activation result. And the corresponding output is input into a linear projector, obtaining $\mathcal{U}_{out}^{i} \in \mathbb{R}^{\rm N \times D}$ of the current operation. Through multiple MPSU-Mamba operations, the decoder output $\mathcal{U}_{out}^{\mathbf{L}} \in \mathbb{R}^{\rm N \times D}$ is taken as input of the bright-channel guided mechanism, which achieves $\mathcal{U}_{d} \in \mathbb{R}^{\rm N \times D}$. Finally, the weighted fusion of $\mathcal{U}_{d}$ and the encoding result $X_{e}$ is taken as the input the final conformation prediction head to obtain the 3D result $\mathbf{G} \in \mathbb{R}^{\rm N \times 3}$. Subsequently, the loss $\mathcal{L}_{decoder}$ is calculated as:
\begin{equation}
\mathcal{L}_{decoder}
= \frac{\displaystyle\sum_{p=1}^{\rm N}\sum_{q=1}^{3}
\left\vert \mathbf{G}[p,q] - \hat{\mathbf{G}}[p,q] \right\vert }{\rm N \times 3}, \quad \mathcal{L}_{opt} = 0.5 * \mathcal{L}_{encoder} + 0.5 * \mathcal{L}_{encoder},
\label{eq8}
\end{equation}
where $\mathcal{L}_{opt}$ represents the final optimized loss. Since the position information is from the encoder, through optimization, the structural understanding of the encoder and decoder could be strengthened effectively.

\subsection{Details of Scanning Strategies}

To construct sufficient understanding of molecular structures, it is important to define proper scanning  strategies.

\textbf{Atom Degree.} In this work, molecules are represented as undirected graphs with multiple nodes and edges. Given an adjacency matrix \({W_{adj}} \in \mathbb{R}^{\rm N \times N} \), the degree of $i$-th atom is calculated as \( \sum_{j=1}^{N} {W_{adj}}[i,j] \). Obviously, atom degree measures the connection tightness between the current atom and other atoms. And the atoms with large degrees may play important roles in the formation of molecular structures.

\textbf{Scanning based on Atom Types.} As shown in Fig. \ref{fig1}a, a molecule could contain multiple different kinds of atoms. To this end, given a molecule, we first statistic the number of each kind of atom. Therefore, in Fig. \ref{fig2}b, for atom types, we simultaneously consider the factors of atom degree and number. The atom corresponding both large degree and number appears earlier in the scanning order. Through this operation, the model could be promoted to focus on critical atom with strong connections, which is beneficial for improving the accuracy of conformation prediction.

\textbf{Scanning based on Connections between Atoms.} In general, besides atoms, various connections usually affect molecular complexity and corresponding molecular characteristics. Though the degree value has considered local connections from neighborhood, long-sequence dependencies may exist molecular structures, still playing an important role. To this end, we employ depth-first search (DFS). Particularly, DFS starts from an initial node and explores as deep as possible along each branch, until no further nodes can be visited, at which point it backtracks and continues to explore other unvisited node.

Concretely, We first initialize a queue to store the atoms with high degrees and those atoms on the ring, then begin the depth-first search from these nodes. Additionally, we also initialize a stack and another queue, which are used to store the atoms to be visited and the current depth-first search path, respectively.

Next, as shown in Fig. \ref{fig2}c and d, an atom from the stack is selected. If this atom has already been visited, DFS skips it and continues with the next atom in the stack. On the contrary, this atom is marked and added to the current path, then visit all of its neighboring atoms. Through performing this process iteratively, we can obtain a global depth-first search path. Obviously, by these operations, this scanning strategy could possess diverse atom relationships including local and long dependencies, which deepens the understanding of molecular structures.

\textbf{Scanning based on Atom Positions.} Essentially, molecule conformation reflects a specific 3D shape. Therefore, it is necessary to consider 3D spatial coordinates. Here, as shown in Fig. \ref{fig3}, we utilize the preliminary predicted molecular conformation \(G_e \in \mathbb{R}^{\rm N \times 3}\). Particularly, along the X, Y, and Z directions, the scanning orders are separately determined based on the coordinate values. Meanwhile, we also employ the forward and backward scanning processes. By this operation, the model could strengthen its understanding of molecular structures. And through optimization based on the loss function Eq. \eqref{eq8}, the prediction accuracy of the encoder could be improved, which achieves superior conformations.

\subsection{Experimental Details}

In the experiments, we first evaluate our MPSU-Mamba method on two existing benchmarks. Meanwhile, to further verify the generalization of our method, we also design a dedicated evaluation on small training data. Experimental results and visualization analysis all demonstrate the effectiveness of our method.

\textbf{Evaluation Metrics.} For molecule's ground-state conformation prediction, we follow the work \cite{kim2025rebind} to utilize four metrics, i.e., D-MAE, D-RMSE, C-RMSD, and E-RMSD. Concretely, given a dataset with $N$ interatomic distances, the Mean Absolute Error (MAE) and Root Mean Square Error (RMSE) between the prediction $\{d_{i}^{*}\}_{i=1}^{N}$ and the ground truth $\{d_{i}\}_{i=1}^{N}$ are used to evaluate the performance at node-pair level:
\begin{equation}
  {\rm D-MAE}(\{d_{i}\}_{i=1}^{N}, \{d_{i}^{*}\}_{i=1}^{N}) = \frac{1}{N} \sum_{i=1}^{N}\left\vert d_{i} - d_{i}^{*} \right\vert,
\end{equation}
\begin{equation}
  {\rm D-RMSE}(\{d_{i}\}_{i=1}^{N}, \{d_{i}^{*}\}_{i=1}^{N}) = \sqrt{\frac{1}{N}\sum_{i=1}^{N}(d_{i} - d_{i}^{*})^{2}}.
\end{equation}

Additionally, C-RMSD score between the ground truth $\hat{\mathbf{G}}$ and the prediction $\mathbf{G}$ is computed as follows:
\begin{equation}
  {\rm C-RMSD}(\hat{\mathbf{G}}, \mathbf{G}) = \sqrt{\frac{1}{n}\sum_{i=1}^{n} {\Vert g_{i}-g_{i}^{*} \Vert}^{2}_{2}}.
\end{equation}

Finally, Energy-weighted RMSD (E-RMSD) is a newly proposed metric by the work \cite{kim2025rebind}, which accounts for the chemical feasibility of predicted conformations:
\begin{equation}
    \text{E-RMSD}(\hat{\mathbf{G}}, \mathbf{G}) = \frac{m}{\widehat{m}}\sqrt{\sum_{i\in\mathcal V}{w}_i\lVert\hat{\mathbf{G}}_i - \mathbf{G}_i\rVert_2},
\end{equation}
where $\frac{m}{\widehat{m}}$ denotes the Boltzmann factor. $w_i$ represents the atom-wise normalized force.

\textbf{Implementation Details. }
Based on the training/validation/test splits of the QM9 and Molecule3D datasets from~\cite{xu2021molecule3d}, we conducted extensive experiments on both datasets. 
% In all experiments, we used a seed of 42, set the embedding dimension to 512, with both the encoder and decoder having 8 layers, a learning rate of 9e-5, a global batch size of 100, and used the AdamW optimizer with no weight decay. 
In all experiments, both the encoder and decoder have 8 layers. We used a seed of 42, set the embedding dimension to 512, the learning rate to 9e-5, the global batch size to 100, and used the AdamW optimizer with no weight decay.
Additionally, for the MPSU-Mamba operation, the hidden dimension was set to 512, the state space dimension was set to 128, the convolution kernel size was set to 4, and the expansion factor was set to 2. All models were trained for 20 epochs, and the best-performing model on the validation set was selected based on the D-MAE metric. All experiments were conducted using PyTorch on RTX 5880 (48GB) and RTX 3090 (24GB) GPU machines. 
Additionally, for the experiments related to GTMGC~\cite{xu2023gtmgc}, REBIND~\cite{kim2025rebind}, and WGFormer~\cite{wang2024wgformer}, we adopted their original settings.

\textbf{Data Availability.} For ground-state conformation prediction, the QM9 dataset can be directly downloaded from the link \url{https://figshare.com/collections/Quantum_chemistry_structures_and_properties_of_134_kilo_molecules/978904}. The Molecule3D dataset can be downloaded from the official website (\url{https://github.com/divelab/MoleculeX/tree/molx/Molecule3D}).

\textbf{Code Availability.} 
For our MPSU-Mamba method, the running codes, including both the training and evaluation scripts, are available at \url{https://github.com/gouyuxin/MPSU-Mamba}. The trained models can also be downloaded from the same repository: \url{https://github.com/gouyuxin/MPSU-Mamba}.


% \bibliographystyle{plain}
\bibliography{sn-bibliography}% common bib file

\end{document}